\documentclass[runningheads]{llncs}

\usepackage[T1]{fontenc}
\usepackage[utf8]{inputenc} 

\usepackage{amsmath,amssymb,bm}

\usepackage{graphicx}
\usepackage{tikz}
\usetikzlibrary{shapes.geometric,arrows}
\usepackage{pgfplots}
\usepackage{pgfplotstable}

\usepackage[table]{xcolor}
\usepackage{array}
\usepackage{multirow}
\usepackage{tabularx}
\usepackage{booktabs}
\usepackage{hhline} 
\usepackage{arydshln} 

\usepackage{listings}
\usepackage[linesnumbered,ruled,vlined]{algorithm2e}
\usepackage{subcaption}     
\usepackage{textcomp}
\usepackage{xspace}
\usepackage{enumitem}
\usepackage{bigdelim}
\usepackage{makecell}
\usepackage{verbatim}
\usepackage{threeparttablex}
\usepackage{calc}
\usepackage{float}
\usepackage{nicematrix}
\usepackage{balance}
\usepackage{cite}
\usepackage{url}
\usepackage{hyperref}
\usepackage{cleveref} 

\definecolor{USTgold}{RGB}{153,102,0}
\definecolor{USTyellow}{RGB}{204,153,0}
\definecolor{USTyellowlight}{RGB}{255,212,0}
\definecolor{USTorange}{RGB}{255,166,26}
\definecolor{USTpink}{RGB}{255,157,157}
\definecolor{USTblue}{RGB}{0,51,102}
\definecolor{USTmiddle}{RGB}{0,116,188}
\definecolor{USTlight}{RGB}{99,202,225}
\definecolor{USTgray}{RGB}{204,204,204}
\definecolor{USTred}{RGB}{237,27,47}
\definecolor{USTdarkred}{RGB}{124,35,72}

\definecolor{CUHKorange}{RGB}{244,106,18}
\definecolor{CUHKblue}{RGB}{0,111,190}
\definecolor{CUHKgreen}{RGB}{0,127,128}
\definecolor{CUHKred}{RGB}{228,46,36}
\definecolor{CUHKyellow}{RGB}{198,148,34}
\definecolor{CUHKdark}{RGB}{114,44,114}
\definecolor{CUHKmiddle}{RGB}{144,44,144}
\definecolor{CUHKlight}{RGB}{167,44,167}
\definecolor{lightblue}{RGB}{223,235,247}

\newcommand{\curly}{\mathrel{\leadsto}}

\newcommand\hoga{\textsc{HOGA}\xspace}
\newcommand\gamora{\textsc{GAmora}\xspace}
\newcommand\deepgate{\textsc{DeepGate}\xspace}
\newcommand{\amulettwo}{\textsc{AMulet2}\xspace}
\newcommand{\revscatwo}{\textsc{RevSCA2}\xspace}
\newcommand{\dynphaseorderopt}{\textsc{DynPhaseOrderOpt}\xspace}
\newcommand{\reveal}{\textsc{ReVEAL}\xspace}
\newcommand{\kissat}{\textsc{Kissat}\xspace}

\newcounter{qst}
\newcommand{\qst}[1]{%
\par\vspace{0.5em}%
\refstepcounter{qst}
\noindent\textbf{\underline{Q\arabic{qst}:}}~\textit{#1}%
\par\vspace{0.5em}%
}

\newcommand{\blackcircled}[1]{%
  \tikz[baseline=(char.base)]{%
    \node[shape=circle,draw,inner sep=1pt,fill=black] (char) {\textcolor{white}{\scriptsize #1}};%
  }%
}

\title{ReVEAL: GNN-Guided Reverse Engineering for Formal Verification of Optimized Multipliers}
\author{
Chen Chen\inst{1} \and
Daniela Kaufmann\inst{2} \and
Chenhui Deng\inst{3} \and
Zhan Song\inst{1} \and
Hongce Zhang\inst{4} \and
Cunxi Yu\inst{1}
}

\institute{
University of Maryland, College Park, USA\\
\email{\{cchen099, zhansong, cunxiyu\}@umd.edu}
\and
TU Wien, Austria\\
\email{daniela.kaufmann@tuwien.ac.at}
\and
NVIDIA\\
\email{cdeng@nvidia.com}
\and
Hong Kong University of Science and Technology (Guangzhou)\\
\email{hongcezh@hkust-gz.edu.cn}
}

\begin{document}
\maketitle

\begin{abstract}
We present \reveal, a graph-learning-based method for reverse engineering of multiplier architectures to improve algebraic circuit verification techniques. Our framework leverages structural graph features and learning-driven inference to identify architecture patterns at scale, enabling robust handling of large optimized multipliers. We demonstrate applicability across diverse multiplier benchmarks and show improvements in scalability and accuracy compared to traditional rule-based approaches. The method integrates smoothly with existing verification flows and supports downstream algebraic proof strategies.
\end{abstract}

\section{Introduction}
\par 
Gate-level arithmetic circuits, in particular integer multipliers that have been optimized via logic synthesis, continue to pose significant challenges for current formal verification techniques. Despite notable advancements in computer algebra-based (CA) techniques~\cite{ciesielski2015verification,kaufmann2023improving,mahzoon2021revsca,konrad2024symbolic}, these methods often rely on syntactic heuristics and hence struggle with synthesized circuits due to optimizations that obscure their original word-level structure \cite{yu2016formal,yu2017fast,mahzoon2021revsca,ciesielski2019understanding,konrad2024symbolic}.

\emph{
Our goal is to verify such optimized circuits by reverse-mapping them back to their original non-optimized word-level representations that can  be efficiently verified using algebraic reasoning. We ensure the correctness of this mapping  through Boolean satisfiability (SAT)-based equivalence checking.}

Traditional computer algebra techniques reduce the task of circuit verification to an ideal membership test for the specification polynomial, typically utilizing Gröbner bases~\cite{Buchberger65} to simplify the polynomials through backward rewriting. In~\cite{KaufmannBiere-TACAS21,kaufmann2023improving}, an incremental column-wise verification approach together with adder substitution is used in the tool \amulettwo to 
handle
unoptimized multipliers. However, the verification of optimized multipliers still presents significant challenges due to  blow-ups in the intermediate rewriting steps. In~\revscatwo\cite{mahzoon2021revsca}
this is addressed by algebraic reverse engineering to recover word-level components such as full adders and half adders. Most recently, \dynphaseorderopt~\cite{konrad2024symbolic} optimizes the encoding of the phases of occurring signals to enhance backward rewriting and keep the sizes of intermediate polynomials small. 
However, despite these advances, logic synthesis often applies aggressive optimizations that obscure the original circuit structure. As a result, many original atomic blocks cannot be restored through cut enumeration, which limits the effect of the aforementioned rewriting methods. 
\begin{figure}[!t]
    \centering
    \includegraphics[width=0.75\textwidth]{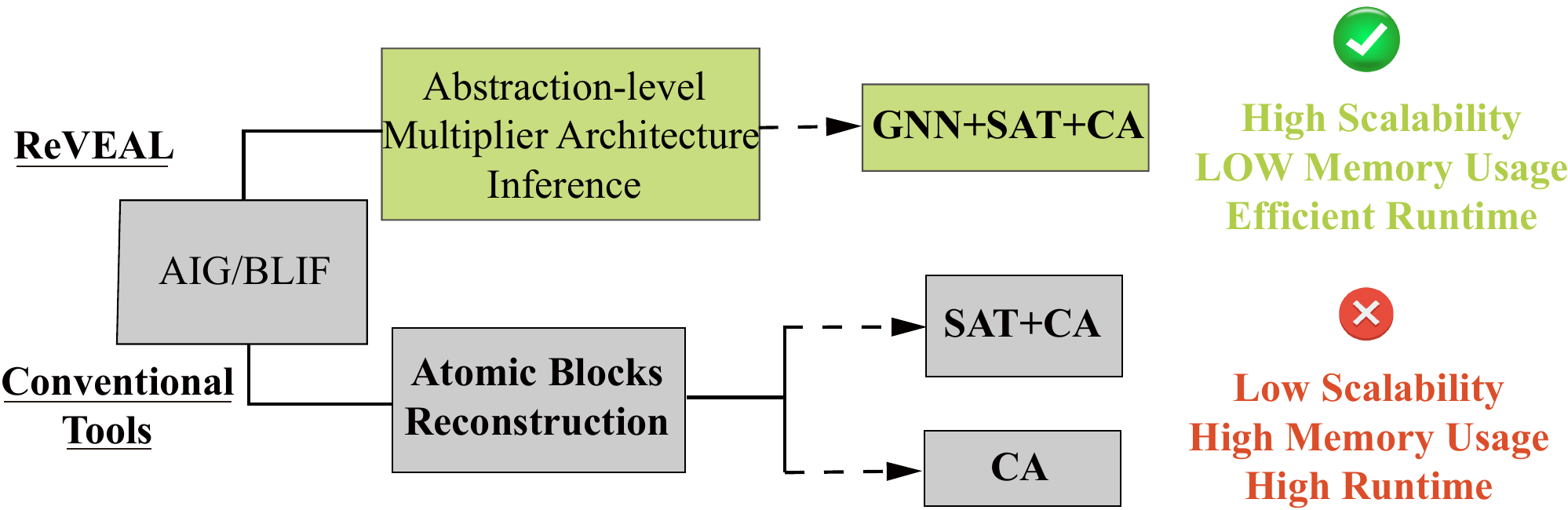}
   
    
    \caption{Overview of ReVEAL versus Conventional Tools} 
    \label{fig:mypdfimage}
\vspace{-3ex}
\end{figure}
Moreover, the inevitable occurrence of numerous OR chains and XOR gates in multiplier circuits further complicates polynomial rewriting, resulting in ongoing issues with monomial blow-ups. 

Recent work by Li et al.~\cite{li2024refscat} reconstructs an adder tree from optimized
multipliers to build a structurally similar reference design, and then verifies it via
CA and SAT-based equivalence checking. However, such reconstruction-based reverse
engineering fundamentally depends on explicit structural pattern matching, which is
fragile under aggressive synthesis and technology mapping. As a result, boundary
recovery can become expensive or fail entirely, and the approach is often
architecture-specific (e.g., tailored to adder trees), making it difficult to handle
diverse designs such as 4-2/compressor and counter-based Wallace trees, XOR-heavy
addition structures, or Booth-encoded multipliers where PPG/PPA boundaries are
blurred.\looseness=-1

Due to the above limitations of reconstruction-based methods~\cite{li2024refscat},
ReVEAL instead formulates the task as a learning-based inference problem.
Traditional reverse engineering largely depends on recovering explicit boundaries and
subgraph patterns, which can be broken or smeared by aggressive synthesis. In contrast, GNNs can aggregate information over neighborhoods
via message passing and encode both local motifs and broader connectivity trends,
allowing ReVEAL to capture more optimization-invariant cues from the netlist
graph and thus remain effective even when canonical structures are not directly
recoverable.

\emph{Our Contributions.} 
In this work, we propose a novel domain-specific GNN-based reverse engineering approach for architectural identification and demonstrate its significant impact in orthogonally accelerating computer algebra-based formal verification techniques by addressing the aforementioned challenges. 

The contributions of this paper are as follows:
\begin{itemize}
    \item 
    
    To the best of our knowledge, ReVEAL is the first framework to integrate machine learning (ML) techniques into an end-to-end formal verification pipeline for multiplier circuits, demonstrating significant improvements in both runtime and scalability.
    
    \item
    We present a novel approach that combines cone extraction with word-level functional and structural features in GNNs. This method enables the model to focus on small graphs within the critical cone, avoiding the performance degradation and long training times typically associated with large graphs. 
    
    \item We compare ReVEAL with advanced (SAT+)CA-based  verification tools. The results show that ReVEAL achieves a 4.90$\times$ speedup and a 19.97$\times$ reduction in memory usage compared to state-of-the-art CA tools. Additionally, ReVEAL outperforms SAT+CA methods with a 13.39$\times$ speedup.


\end{itemize}

\section{Preliminaries}
\vspace{-1ex}
In this section we provide background information on different multiplier architectures (Section~\ref{sec:mult}), a short introduction to GNNs (Section~\ref{sec:gnn}), as well as a very brief discussion of SAT solvers and SAT sweeping (Section~\ref{sec:SAT}).


\vspace{-1ex}
\subsection{Multiplier Architectures}\label{sec:mult}
Multipliers typically involve three stages: (1) \emph{Partial Product Generator (PPG)}, (2) \emph{Partial Product Accumulator (PPA)}, and (3) \emph{Final Stage Adder~(FSA)}. 

The PPG can employ either a simple approach that pairs corresponding bits of the  multiplicands to generate each partial product using logical conjunction or a Booth encoding scheme to process multiple bits at a time. 

The PPA then sums these partial products using structures like Wallace trees, Dadda trees, or 4-to-2 compressor trees, which reduce gate delays and minimize sequential additions compared to traditional array accumulators.

Finally, the FSA combines the last two rows of partial products using adders, which
can be broadly categorized into non-tree-based and tree-based adders (e.g.,
Brent--Kung, Kogge--Stone, Sklansky, Han--Carlson), with tree adders typically
showing more regular, highly connected prefix structures. We list the most common
multiplier architectures in \Cref{tab:arc_adders}.
 \begin{table}[tb]
  \centering
  \caption{Categorization of Multiplier Architectures }
  \resizebox{\columnwidth}{!}{
    \begin{tabular}{c|p{0.45\columnwidth}p{0.45\columnwidth}}
      \hline
      \textbf{Stage} & \multicolumn{2}{c}{\textbf{Types}} \\
      \hline
      
      \multirow{1}{*}{\blackcircled{1} PPG}
      & $\bullet$ Booth Encoding
      & $\bullet$ Simple  \\
      
      \hline
      
      \multirow{3}{*}{\blackcircled{2} PPA}
      & $\bullet$ Simple Array
      & $\bullet$ Wallace Tree \\
      & $\bullet$ Dadda Tree
      & $\bullet$ 4-to-2 Compressor Tree \\
      & $\bullet$ Counter-based Wallace Tree
      & \\
      \hline
      
      \multirow{6}{*}{\blackcircled{3} FSA}
      & $\bullet$ Ripple Carry Adder
      & $\bullet$ Carry Look-Ahead Adder \\
      & $\bullet$ Carry Skip Adder
      & $\bullet$ Serial Prefix Adder \\
      & \cellcolor{lightblue}$\bullet$ Brent-Kung Adder
      & \cellcolor{lightblue}$\bullet$ Sklansky Adder \\
      & \cellcolor{lightblue}$\bullet$ Han-Carlson Adder
      & \cellcolor{lightblue}$\bullet$ Ladner-Fischer Adder \\
      & \cellcolor{lightblue}$\bullet$ Kogge-Stone Adder 
      & \\
      \hline
    \end{tabular}
  }
  \textbf{Note:} Tree adders are marked using blue cell colors.

   \label{tab:arc_adders}
  \vspace{-3ex}
\end{table}

\subsection{Graph Neural Networks}\label{sec:gnn}

We refer readers to~\cite{hamilton2020graph} for a comprehensive introduction to graph representation learning and GNNs.
A GNN encodes a graph \( G = (V, E) \) into a more compact representation.  Starting with initial node features \( \{ \mathbf{x}^{(0)}_i \mid i \in V \} \), GNNs iteratively update hidden state vectors $\mathbf{x}^{(l)}_i$ using:

\begin{align}
\mathbf{m}^{(l)}_i &= \text{AGGREGATE}^{(l)} \left( \{ \mathbf{x}^{(l-1)}_j \mid j \in \mathcal{N}(i) \} \right), \label{eq:aggregate} \\
\mathbf{x}^{(l)}_i &= \text{COMBINE}^{(l)} \left( \mathbf{m}^{(l)}_i, \mathbf{x}^{(l-1)}_i \right), \label{eq:combine}
\end{align}

where \( \mathcal{N}(i) \) are the one-hop neighbors of node \( i \). The AGGREGATE function (e.g., mean or max pooling) pools neighbor features, and the COMBINE function integrates this with the node's current feature.
After \( L \) layers, 
the set of vectors
\( \{ \mathbf{x}^{(L)}_i \mid i \in V \} \) capture the structural and feature information of the graph. A readout function then aggregates these into a graph embedding for classification, enabling GNNs to effectively model complex dependencies in circuit designs for various EDA applications.

Building on this capability, recent research has explored the use of GNNs for reverse engineering tasks. \hoga~\cite{deng2024less} and \gamora~\cite{wu2023gamora} leverage gate-level Boolean reasoning to identify adders at the register transfer level (RTL), while the \deepgate family~\cite{shi2024deepgate3} utilizes Boolean circuit representation learning to facilitate SAT solving. Furthermore, FGNN~\cite{wang2024fgnn2,he2021graph} employs graph learning techniques to identify adder blocks and classify circuits. 

These prior works studied the use of GNNs in low-level atomic block identification/reconstruction, while our proposed approach ReVEAL is the first work that investigates the application of GNNs in the inference of high-level design choices (abstraction-level architectural inference).

\subsection{Modern SAT Solvers and SAT Sweeping}\label{sec:SAT}
\kissat~\cite{biere2024clausal1,biere2024clausal} is a state-of-the-art SAT solver, which employs clausal congruence closure 
to efficiently handle large-scale equivalence checking problems, particularly for circuits containing numerous small isomorphic subcircuits.

SAT sweeping~\cite{DBLP:conf/dac/ZhangJAMB21} is an efficient equivalence checking approach that combines simulation with incremental SAT solving. It detects equivalent nodes through simulation, constructs intermediate miters, and runs SAT solving in topological order on two circuits to iteratively refine and merge equivalent nodes.

\subsection{Problem Formulation}
\label{sec:problem_formulation}
We denote an optimized multiplier gate-level netlist in AIG form as $G$ with
bit-width $N$. Table~\ref{tab:arc_adders} summarizes the candidate structures
for the three stages considered in this work. Given $G$, \textsc{ReVEAL} infers
the corresponding (PPG, PPA, FSA) architecture choices from
Table~\ref{tab:arc_adders}. Based on the inferred architectures and $N$, we select the
matching pre-verified template from our library and use SAT-based
combinational equivalence checking (CEC) to validate functional equivalence
between $G$ and the selected template.

\section{Methodology -- the ReVEAL Framework}\label{sec:our-method}

\begin{figure*}[tb]

\centering
\includegraphics[width=\textwidth, height=0.45\textwidth]{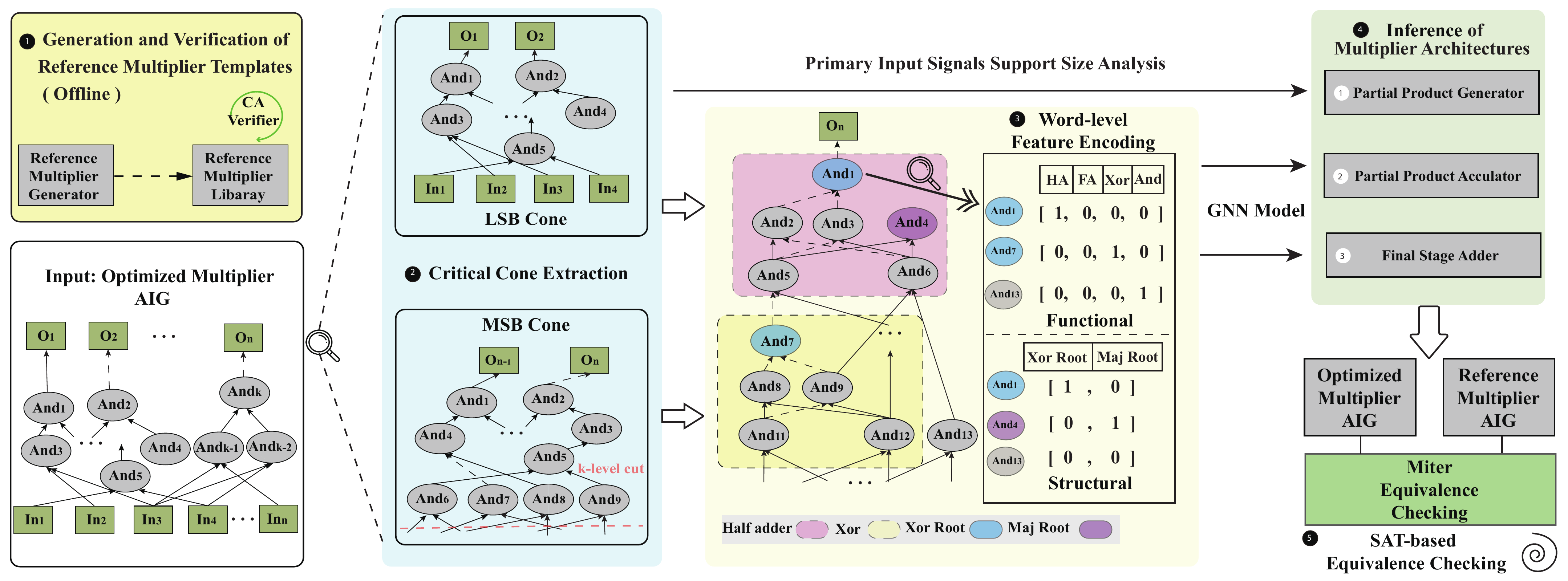}
\caption{The Workflow of ReVEAL}
\label{fig:overview}

\end{figure*}
\Cref{fig:overview} presents the workflow of our reverse engineering framework \textbf{ReVEAL}, which consists of five main steps. Steps 2--4 are explained in more detail in the following sections.

\textbf{\blackcircled{1} Offline Generation and Verification of Reference Multiplier Templates.}  
We generate RTL templates for multipliers of various bitwidths and architectures using multiplier generators~\cite{Mahzoon2021GenMulGA}~\cite{DBLP:conf/cav/TemelSH20}. These templates are processed with Yosys~\cite{yosys} to produce their corresponding And-Inverter Graphs (AIGs).
All templates are formally certified using \amulettwo~\cite{kaufmann2023improving} and are collected in our \emph{reference multiplier library}.
Notably, the generation and certification processes are performed offline.

\textbf{\blackcircled{2} Critical Cone Extraction.}  
The input to ReVEAL is a gate-level multiplier \( \mathcal{G} \), optimized by logic synthesis tools, such as ABC~\cite{DBLP:conf/dac/MishchenkoCB06}. We extract the Least Significant Bit (LSB) and Most Significant Bit (MSB) cones of \( \mathcal{G} \), which allows us to perform architecture inference on smaller subcircuits and hence enhances the generalization capability of the GNN model.

\textbf{\blackcircled{3} Word-level Feature Encoding.}  
For the extracted LSB and MSB cones, we perform reverse identification of arithmetic blocks such as full adders and half adders to encode word-level features on \( \mathcal{G} \).

\textbf{\blackcircled{4} Inference of Multiplier Architecture.}  
By combining graph analysis on primary input signals and GNN model inference, we predict the architecture of the optimized multiplier, identifying its PPG, PPA, FSA.

\textbf{\blackcircled{5} SAT-based Equivalence Checking.} 
Using the generated reference multiplier library of step 1, we locate a pre-verified template with the same architecture as the prediction of the optimized multiplier. A miter circuit is constructed by combining the optimized multiplier and the reference template for equivalence checking. We verify the miter using SAT solving/sweeping; see \textbf{Q\ref{exp_solver}} in the experimental evaluation.




\subsection{Critical Cone Extraction}\label{sec:cone}
Previously, GNNs struggled with long-range dependencies in large-scale graphs, leading to over-smoothing~\cite{DBLP:conf/aaai/LiHW18} and over-squashing~\cite{DBLP:conf/iclr/0002Y21}. Over-smoothing makes node representations indistinct, while over-squashing compresses excessive information into fixed-size embeddings, causing loss of critical details. These issues hinder the performance and scalability of GNNs on large graph structures. 

\begin{figure}[tb]
    \centering    \includegraphics[width=1\textwidth ]{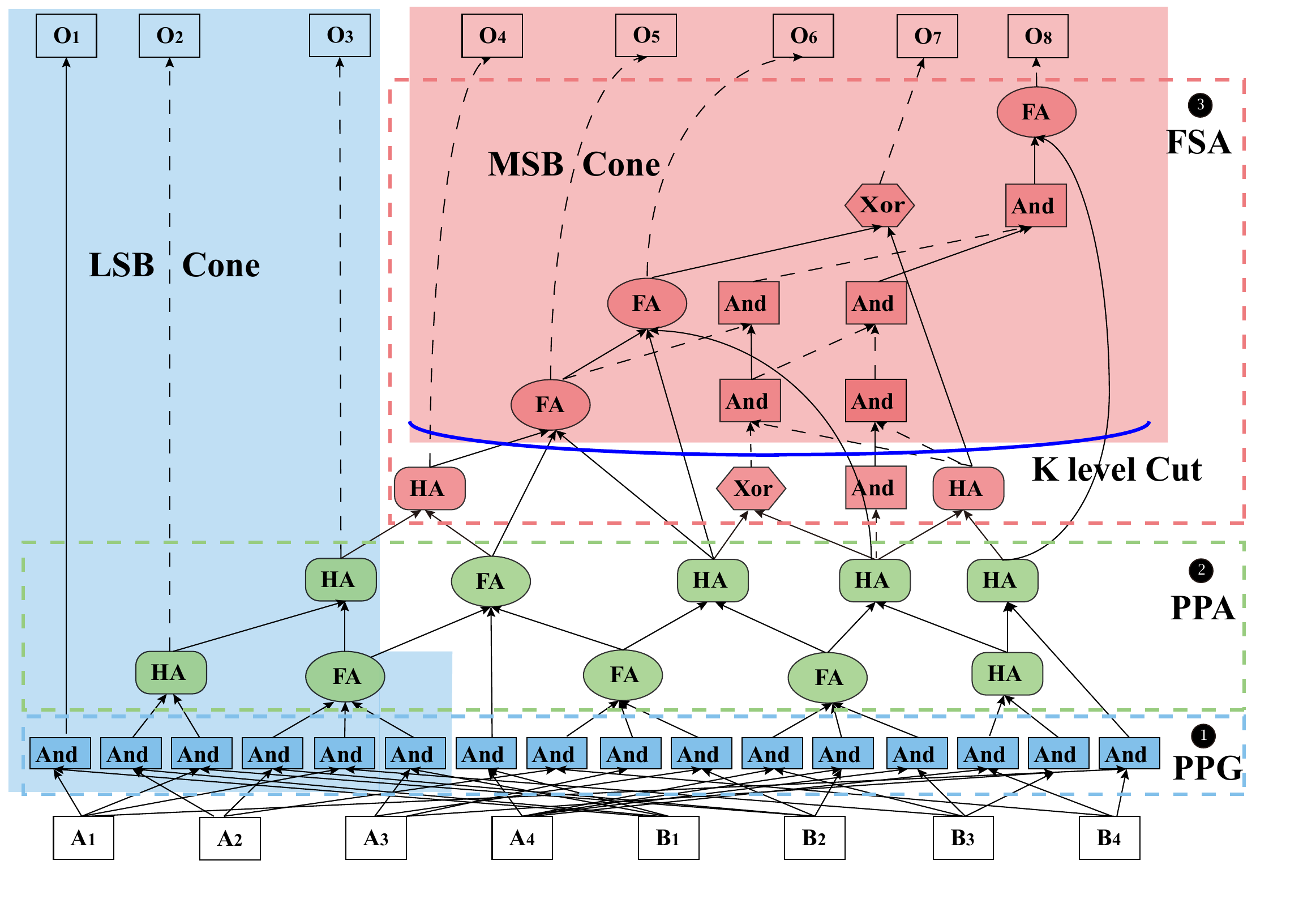} 
    \caption{Example of LSB \& MSB cone extraction for a 4-bit multiplier }
    \label{fig:runtime_scalability}
    \vspace{-3ex}
\end{figure}

To tackle these challenges, we partition the graph to concentrate on critical cones relevant to their corresponding primary stages of the multiplier architecture: PPG, PPA, FSA, see \Cref{fig:runtime_scalability} for an example. 
Firstly we handle the \textbf{\blackcircled{1}} PPG and \textbf{\blackcircled{2}} PPA stages by pre-processing the input AIG and extract the 
\(C\) least significant primary outputs and their corresponding input cones. In our experimental dataset (see \Cref{sec:experiment}), which starts with a minimum input bit-width of 32, we have empirically observed that the cones of the \(C\) = 8 LSBs contain sufficient distinct information to successfully identify the \textbf{\blackcircled{2}} PPA, without being affected by contamination
 from the \textbf{\blackcircled{3}} FSA stages. Moreover, as the bit-width increases, the cones of the first 8 LSBs trace back to 16 input bits, displaying highly similar cone structures, which are beneficial for GNNs generalization.

Regarding the \textbf{\blackcircled{1}} PPG, we directly perform graph analysis on the cone extracted from the LSBs. 
For doing so, we define the support size $\text{SuppSize}(X)$ for a node $X$ as:
\hspace{-2cm}
\[
\text{SuppSize}(X) = \left| \{ I \in \{\text{Primary Inputs}\} : \exists \, I \curly X \} \right|,
\]
%
 where $\exists \, I \curly X$
indicates that there exists a directed path from primary input node $I$ to node $X$. It recursively traverses its fan-in nodes, incrementally counting the primary inputs that influence the outputs.

We extract a graph slice where the support size of the primary input signals is~2. 
Since Booth multipliers involve multiple input signals for each product generator, by analyzing the fanout of primary inputs and the logic levels within the slice, we can determine the use of Booth encoding. Specifically, when the supporting size is set to 2 and the slice has more than 2 logic levels, it indicates that Booth encoding is being utilized. Similarly, we can also identify higher-radix multipliers
 by examining the output support nodes of the primary inputs, as Booth-$n$ encoding, $n+1$ bits of the multiplier are examined, and $n$ multiplier bits are encoded.\looseness=-1

For the \textbf{\blackcircled{3}}  FSA stage, we begin at the MSB, denoted as \( M \), and extract the cone spanning from the \( M \)-th bit to the \( (M - L) \)-th bit, where \( L = 8 \).
This value is again empirically derived.
To further isolate this cone from potential interference by the PPA and PPG stages, we implement a \( K \)-level cut of the MSB cone. The selection of \( K \) is contingent upon the multiplier bit-width~\( N \). Specifically, through empirical observation and symbolic regression on our experimental dataset, we have established a relationship between \( N \) and the corresponding logic level \( K \) required for effective graph partitioning; that is:
$
K = 5 + 2 \cdot \left\lfloor \log_2(N) \right\rfloor
$.
This formula ensures that for varying bit-widths—from 32 to 256 bits—the \( K \)-level cut adapts appropriately, maintaining consistent cone sizes and effectively focusing only on small cones relevant to the FSA stages. Combining the dynamic \( K \)-level cut with a fixed \( L = 8 \) ensures that the extracted cone retains critical structural and functional information pertinent to the FSA, while keeping the cone size within a manageable limit of a few hundred nodes. This size restriction is crucial for enhancing the training speed, allowing the learning process to remain efficient and maintain generalization during inference at larger bit widths.
\begin{table}[t]
  \centering
   \caption{Feature Embeddings for Graph Neural Network Training}
  \label{tab:one_hot_encoding}
  \resizebox{0.6\columnwidth}{!}{
  \begin{tabular}{c|c|l|c}
    \hline
    \textbf{Level} & \textbf{Type} & \textbf{Feature} & \textbf{Value Type} \\
    \hline
    \hline
    \multirow{7}{*}{\textbf{Node}} 
    & \multirow{3}{*}{Basic} 
    & $\bullet$ Input edge1 inversion  & 0/1 \\
    & & $\bullet$ Input edge2 inversion  & 0/1 \\
     && $\bullet$ PI/PO/inner node & 0/1 \\
    \cline{2-4}
    & \multirow{4}{*}{Function}
    & $\bullet$ HA node & 0/1 \\
    & & $\bullet$ FA node & 0/1 \\
    & & $\bullet$ Remaining XOR node & 0/1 \\
    & & $\bullet$ AND node & 0/1 \\
    \cline{2-4}
    & \multirow{2}{*}{Structure}
    &  $\bullet$ MAJ root & 0/1 \\
    & & $\bullet$ XOR root & 0/1 \\
    \hline
    \multirow{4}{*}{\textbf{Graph}} 
    & \multirow{4}{*}{Structure}
    & $\bullet$ Input count & Integer \\
    & & $\bullet$ Gate count & Integer \\
    & & $\bullet$ Graph density & Float \\
    & & $\bullet$ Clustering Coefficient & Float \\
    \hline
  \end{tabular}
  \label{tab:feature_sum}
  }
  \vspace{-2em}
\end{table}

\subsection{Word-Level Feature Encoding} 

Feature representation is key to effective message propagation in GNNs, as it determines how information is transmitted and aggregated across the graph. 

\Cref{tab:feature_sum} summarizes our exploration of node-level and graph-level features. In addition to retaining the basic structural feature encoding used in \hoga and \gamora, which employs a one-hot feature vector to represent node types such as primary inputs (PI), internal nodes, and primary outputs (PO), as well as inversions on edge, we enhance basic functional node features to enable reverse mapping to word-level atomic blocks~\cite{DBLP:conf/iscas/YuC16}.  

For functional atomic block identification, we employ cut enumeration~\cite{DBLP:conf/fpga/PanL98} in the graph~\( \mathcal{G} \) to reverse-engineer the functional components at the RTL level, identifying elements such as full adders (FA) and half adders (HA), and remaining XOR gates that are not part of FA/HAs. For readers interested in the details of identification, we provide the complete procedure in~\Cref{sec:word_level_id}.
By detecting all XOR3 and MAJ3\footnote{XOR3 refers to the logic gate implementing the three-input parity function~$x \oplus y \oplus z$; MAJ3 denotes the three-input majority function~$\text{maj}(x, y, z) = (x \wedge y) \vee (x \wedge z) \vee (y \wedge z)$.} combinations in the FAs, we pinpoint the corresponding input nodes (\(\mathit{IN}_1\), \(\mathit{IN}_2\), \(\mathit{IN}_3\)) and output nodes (\textit{XOR3 Root}, \textit{MAJ3 Root}). The same method applies to HAs. Through recursive traversal of the input and output nodes in each pair, we identify all AIG nodes
that can be reverse-mapped to HAs and FAs. The remaining nodes that can be reverse-mapped to an XOR are annotated as XOR nodes, while the remaining AIG nodes are annotated as AND gates. In the examples shown in \Cref{fig:overview}, nodes such as AND1, AND7, and AND13 are each represented by a  vector that indicates their respective functional blocks as HAs, FAs, XOR, and AND.

Cone extraction, as discussed in~\Cref{sec:cone}, enables us to work with smaller graphs, allowing for the quick computation of graph-level features such as graph density\footnote{%
Graph density: fraction of present edges out of all possible (0–1; higher means more connections).}, clustering coefficient\footnote{
Clustering coefficient: how tightly a node’s neighbors connect to each other (higher means a tighter group).}, cut input count, and gate count.

These features are crucial for identifying the sparsity of connections and cut fan-in numbers in different adder structures. 
Additionally, the time required for generating these functional and structural node features is negligible.
This efficiency in feature computation significantly enhances our model's ability to classify different types of adders while maintaining scalability in larger designs.


\subsection{Inference of Multiplier Architecture} 
\label{sec:framework}

We use a multi-task hierarchical classification strategy built upon a bidirectional GraphSAGE~\cite{DBLP:conf/iccad/UstunDP0Z20} architecture to classify and more accurately distinguish different architectures of PPAs and FSAs. Recall, that we only require inference for PPA and FSA, as PPG can already be classified during critical cone extraction.

\subsubsection{Directional GraphSAGE Model.}
The model captures directional dependencies through parallel message passing, where the graph neural network aggregates features from both incoming (predecessor) and outgoing (successor) edges. The directional embedding computation at layer $k$ is defined as:

\begin{equation}
\begin{aligned}
\mathbf{h}_v^{\text{pre}(k)} &= \text{ReLU}\left(\mathbf{W}_{\text{pre}}^{(k)} \cdot \text{MEAN}\left(\{\mathbf{h}_u^{(k-1)} \mid u \in \mathcal{N}_{\text{in}}(v)\}\right)\right) \\
\mathbf{h}_v^{\text{suc}(k)} &= \text{ReLU}\left(\mathbf{W}_{\text{suc}}^{(k)} \cdot \text{MEAN}\left(\{\mathbf{h}_u^{(k-1)} \mid u \in \mathcal{N}_{\text{out}}(v)\}\right)\right)
\end{aligned}
\label{eq:bidirectional}
\end{equation}

with $\mathbf{W}_{\text{pre}}^{(k)}$ and $\mathbf{W}_{\text{suc}}^{(k)}$ being directional message parsing weights for layer $k$, and where $ \mathcal{N}_{\text{in}}(v)$ resp. $\mathcal{N}_{\text{out}}(v)$ are the one-hop neighbors of node \( v \)

After three message-passing layers, we compute normalized node embeddings through concatenation ($\parallel$) and batch normalization (BN) as follows:

\begin{equation}
\mathbf{h}_v^{\text{final}} = \text{BN}\left(\mathbf{h}_v^{\text{pre}(3)} \parallel \mathbf{h}_v^{\text{suc}(3)}\right) \quad \forall v \in G
\label{eq:node_embed}
\end{equation}
and then obtain the graph-level representation via global mean pooling:
\begin{equation}
\mathbf{h}_G = \frac{1}{|G|} \sum_{v \in G} \mathbf{h}_v^{\text{final}}
\label{eq:graph_pool}
\end{equation}

\subsubsection{Hierarchical Classification.}
Our two-stage hierarchical design addresses the error accumulation problem in a direct 9-class FSA classification (cf.~\Cref{tab:arc_adders}) through structural decomposition. The key insight stems from the fundamental architectural differences between tree-based and non-tree-based adders: tree structures exhibit regular logarithmic-depth patterns, while non-tree implementations often contain irregular subgraph motifs. This hierarchical strategy first performs a coarse-grained binary classification to distinguish tree-based and non-tree-based adders, followed by fine-grained classification within each category to identify specific adder types. The classification pipeline proceeds as follows:

\paragraph{Stage 1 - Tree vs. Non-tree Discrimination}
\begin{equation}
\hat{y}_{\text{topo}} = \arg\max\left(\mathbf{W}_{\text{topo}}[\mathbf{h}_G \parallel f_{\text{level}}]\right)
\end{equation}
where $f_{\text{level}}$ denotes the maximum logic depth of the optimized circuit, capturing critical structural information. This stage effectively separates tree adders, characterized by deep recursive structures, from non-tree adders, which exhibit shallow combinational designs.

\vspace{1em}
\paragraph{Stage 2 - Fine-grained Classification}
Conditioned on the predicted topology (tree vs.\ non-tree), we perform subtype
classification by combining the graph embedding with lightweight structural
summaries:
\begin{equation}
\hat{y}_{\text{sub}} =
\arg\max\!\left(\mathbf{W}_{t}\,[\mathbf{h}_G \parallel f_{\text{fan}} \parallel \mathbf{f}_{\text{graph}}]\right),
\qquad t \in \{\text{nt},\text{tree}\}.
\end{equation}
Here, $f_{\text{fan}}$ is the fan-in statistic extracted from the $K$-level MSB cone, and
$\mathbf{f}_{\text{graph}}\in\mathbb{R}^3$ summarizes global structure (density, clustering
coefficient, and average degree).

\subsubsection{Multi-task Training Strategy.}
\label{sec:training}
The model's trainable parameters $\Theta$ comprise three components: directional message passing weights $\mathbf{W}_{\text{pre}}^{(k)}$ and $\mathbf{W}_{\text{suc}}^{(k)}$ from Eq.~\ref{eq:bidirectional}, where $k \in \{1, 2, 3\}$ represents the three layers of message passing, along with classifier weights $\mathbf{W}_{\text{topo}}$ for topology prediction, $\mathbf{W}_{\text{nt}}$ for non-tree subtypes, and $\mathbf{W}_{\text{tree}}$ for tree subtypes. The gradient composition rule combines task-specific gradients through conditional routing:
\begin{equation}
\frac{\partial\mathcal{L}}{\partial\Theta} = \frac{\partial\mathcal{L}_{\text{topo}}}{\partial\Theta} + \alpha\left[ I(\hat{y}_{\text{topo}}=0)\frac{\partial\mathcal{L}_{\text{nt}}}{\partial\Theta} + I(\hat{y}_{\text{topo}}=1)\frac{\partial\mathcal{L}_{\text{tree}}}{\partial\Theta} \right],
\end{equation}
where $I(\cdot)$ implements conditional gradient routing based on topology prediction $\hat{y}_{\text{topo}}$, with $\mathcal{L}_{\text{topo}}$ denoting binary cross-entropy loss, $\mathcal{L}_{\text{nt}}$ for 4-class non-tree classification, and $\mathcal{L}_{\text{tree}}$ for 5-class tree classification.
The balancing coefficient $\alpha=3.2$ is calibrated during initial warm-up training through gradient variance analysis. This process computes the Frobenius norms' squared magnitudes $\|\nabla_\Theta\mathcal{L}_{\text{topo}}\|_F^2$ and $\|\nabla_\Theta\mathcal{L}_{\text{sub}}\|_F^2$ over 100 iterations, where $\mathcal{L}_{\text{sub}} = \mathcal{L}_{\text{nt}} + \mathcal{L}_{\text{tree}}$. Statistical aggregation calculates the running averages $\mathbb{E}[\cdot] = \frac{1}{100}\sum_{t=1}^{100}(\cdot)_t$, determining the optimal coefficient $\alpha$ via:
\begin{equation}
\alpha = \sqrt{\frac{\mathbb{E}[\|\nabla_\Theta\mathcal{L}_{\text{topo}}\|_F^2]}{\mathbb{E}[\|\nabla_\Theta\mathcal{L}_{\text{sub}}\|_F^2]}} \approx 3.2,
\end{equation}
where $\|\cdot\|_F$ denotes the Frobenius norm, and $\|\cdot\|_F^2$ represents its squared magnitude, which is used here to measure the gradient strength for balancing tasks.

Parameter updates follow distinct pathways: message passing weights $\mathbf{W}_{\text{pre/suc}}$ receive combined gradients from both tasks, while $\mathbf{W}_{\text{topo}}$ updates exclusively through $\mathcal{L}_{\text{topo}}$. The subtype classifiers $\mathbf{W}_{\text{nt}}$ and $\mathbf{W}_{\text{tree}}$ activate only when $\hat{y}_{\text{topo}}$ predicts non-tree (0) or tree (1) topologies, respectively. The unified update rule preserves batch-normalized stability (Eq.~\ref{eq:node_embed}):
\begin{equation}
\Theta^{(t+1)} = \Theta^{(t)} - \eta\left(\nabla_\Theta\mathcal{L}_{\text{topo}} + \alpha\nabla_\Theta\mathcal{L}_{\text{sub}}\right).
\end{equation}
The proposed hierarchical classification framework, coupled with the multi-task training strategy, effectively mitigates error accumulation and ensures robust performance by leveraging topology-aware feature decomposition and gradient-balanced optimization.

\section{Experiments}\label{sec:experiment}
\paragraph{Experimental Setup.}\label{sec:exp-setup}
All experiments are run on a server running Ubuntu 20.04.4 LTS, with Intel(R) Xeon(R) Gold 6418H processors and 64 GB of memory.
We set a timeout (TO) of 3600~seconds and a memory limit (MO) of 8~GB.

All multiplier circuits are converted to AIG format using Yosys~\cite{yosys} and ABC~\cite{DBLP:conf/dac/MishchenkoCB06}. 
All generated reference multipliers are verified using \amulettwo~\cite{kaufmann2023improving} with 64 pre-prepared templates per bit-width (step 1). 
The algorithms for critical cone extraction (step 2) and word-level feature encoding (step 3) are implemented and integrated in ABC. 
For miter equivalence checking (step 5) we consider the SAT solver \kissat 4.0.1~\cite{biere2024clausal} and ABC’s SAT-sweeping-based combinational equivalence checkers \texttt{\&fraig}, \texttt{cec} and \texttt{\&cec}; the rationale for this solver combination is analyzed in \textbf{Q\ref{exp_solver}}. 

\paragraph{Dataset.}
We use the multipliers generated by the generators Multgen~\cite{DBLP:conf/cav/TemelSH20} and GenMul~\cite{Mahzoon2021GenMulGA}\footnote{We do not consider the richer Aoki benchmark set~\cite{aokipaper} in our analysis due to the discontinuation of their online generator, which renders it impossible to obtain the small-width cases necessary for model training.}. GenMul provides 28 distinct architectures, while Multgen supports 36 architectures. We include all available architectures with one exception: for Multgen, we exclude Booth-encoding templates at bit-widths of 4 and above, because these instances cannot be correctly parsed by Yosys during SystemVerilog-to-AIG conversion. 

To build the training and evaluation sets, we use even-numbered bit-widths from 32 to 60 bits for training and evaluate on unseen 64-, 128-, and 256-bit designs. Following prior work~\cite{mahzoon2021revsca,konrad2024symbolic}, all optimized multipliers are produced using ABC operators: \texttt{dc2} or \texttt{resyn3}. All selected designs—both original and optimized—are normalized to AIG using Yosys~\cite{yosys} and ABC~\cite{DBLP:conf/dac/MishchenkoCB06}, ensuring consistent representation across generators, optimizations, and bit-widths. This design enables a rigorous assessment of the model’s ability to generalize to unseen, larger bit-widths, beyond the training regimes.

\paragraph{Machine Learning Configuration.}
\label{exp_setting}
We train our model for 200 epochs per optimization operator. All hyperparameters
(e.g., learning rate, batch size, and weight decay) are automatically tuned using
Optuna~\cite{akiba2019optuna}; the final configurations and trained model checkpoints
are provided in our open-source repository. Our pipeline demonstrates exceptional efficiency: cone extraction plus word-level feature extraction complete in under 1 second per case. A complete training cycle consists of 329 seconds for feature extraction and 321 seconds for training, totaling 650 seconds. 
Our training scales well as bit-widths grow thanks to critical cone extraction.
In contrast, conventional full-graph training approaches, which perform equivalent computations on full AIGs without cone extraction, are 182× slower.\looseness=-1

\ \\
\textbf{Research Questions.}
We organize the presentation of experimental results to address the following research questions:

\qst{Can GNNs and analytical techniques demonstrate robust performance and generalization capability for architectural inference across optimized netlists?} \label{exp:preprocess}

\begin{table}[tb]
   \vspace{-1em} 
    \centering
    \caption{Inference Accuracy for Each Stage}
    \begin{tabular}{lccc}
        \toprule
        \textbf{Stage} & \textbf{TOP 1}$^*$ & \textbf{TOP 2}$^*$ & \textbf{TOP 3}$^*$ \\
        \midrule
        Stage PPG & 100\% & -- & -- \\
        Stage PPA & 100\% & --  & -- \\
        Stage FSA & 85\% & 97\% & 98\% \\
        \bottomrule
    \end{tabular}
    \label{tab:stage_accuracy}

    {\raggedright
    \footnotesize
    $^*$ TOP 1, TOP 2, and TOP 3 represent the prediction accuracy when the ground truth is among the model's top 1, top 2, and top 3 ranked predictions, respectively. \par
    }
\end{table}

 The inference accuracy is shown in \Cref{tab:stage_accuracy}. We also summarize, for the 64-bit dc2-optimized case, all top-3 predictions in \Cref{tab:dc2-64-gnn-rank-part2}.
 In \emph{Stage PPA} we observe that the LSBs cone maintains similar cone graph sizes and structural similarities across different bit-widths, enhancing the model’s learning capability for larger bit-widths. As a result, our training process achieves an accuracy rate of 100\% for the first two stages. Furthermore, our model demonstrates high accuracy in \emph{Stage FSA}, with the top three predictions nearly achieving complete correctness. 
 We observe that when the circuits have been optimized using \texttt{dc2}, the FSA prediction errors mainly appear on Brent–Kung (BK) cases. This happens because our MSB-centered cone extraction is only an approximate partition. 
 In some extensively optimized cases, we observe that structurally similar but not identical templates of the FSA can facilitate faster equivalence checking. Therefore, we retain the top three predicted templates and have them participate in SAT solving in parallel.







\qst{Are modern SAT solvers and SAT sweeping techniques efficient for equivalence checking between the optimized and the golden multiplier?}\label{exp_solver}

\begin{table}[tb]
\centering
\caption{Solver Performance for Miter Solving}
\label{tab:solver_performance}
\resizebox{0.8\textwidth}{!}{%
\begin{tabular}{@{}lcccccc@{}}
\toprule
\textbf{Opt}        & \multicolumn{3}{c}{\textbf{dc2}} & \multicolumn{3}{c}{\textbf{resyn3}} \\ \cmidrule(lr){2-4} \cmidrule(lr){5-7}
                    & \textbf{Solved} & \textbf{Avg Time (s)} & \textbf{\#. fastest} & \textbf{Solved} & \textbf{Avg Time (s)} & \textbf{\#. fastest}\\ \midrule
\centering cec      & 60/64           & 692.57                & 0 & 60/64           & 1007.20               & 1 \\
\centering \&cec    & 64/64           & 8.50                  & 46& 64/64           & 6.83                 & 50\\
\centering \&fraig -y & 36/64          & 1848.53               & 12& 39/64           & 1877.33              & 8 \\
\centering \&fraig -x & 37/64          & 1765.14               & 0& 39/64           & 1726.71              &  5 \\
\centering kissat   & 64/64           & 96.79                 & 6& 64/64           & 186.55               & 0 \\ \bottomrule
\end{tabular}%
}
\end{table}

We selected the SAT-sweeping tools mentioned in the experimental setup for equivalence checking of optimized (synthesized) and original (golden) circuits. 
%
%
%
We tested 64-bit multipliers with 64 different templates. 
Our quantitative analysis in Table~\ref{tab:solver_performance} reveals that while \texttt{\&cec} achieves the highest success rate, \kissat, \texttt{\&fraig -y} and \texttt{\&cec} remain essential for handling specific corner cases. Notably, the SAT-sweeping approach demonstrates remarkable memory efficiency compared to computer algebra-based methods --- a critical advantage when verifying complex arithmetic circuits, which  we will quantitatively validate in terms of memory consumption later for \textbf{Q\ref{exp:rq2}}. Based on these observations, we implement a hybrid parallel verification framework that concurrently executes three strategically selected solvers (\texttt{\&fraig -y}, \texttt{\&cec}, and \kissat), terminating the verification process immediately when any solver returns a conclusive result.


%


\qst{Is architecture identification helpful for computer-algebra-based multiplier verification?}\label{exp:rq2}
\vspace{-3ex}

\paragraph{Comparison with State-of-the-Art CA Verification Tools.}

\begin{figure*}[t]
    \centering
    \begin{subfigure}[b]{0.24\linewidth}
        \centering
        \includegraphics[width=\textwidth]{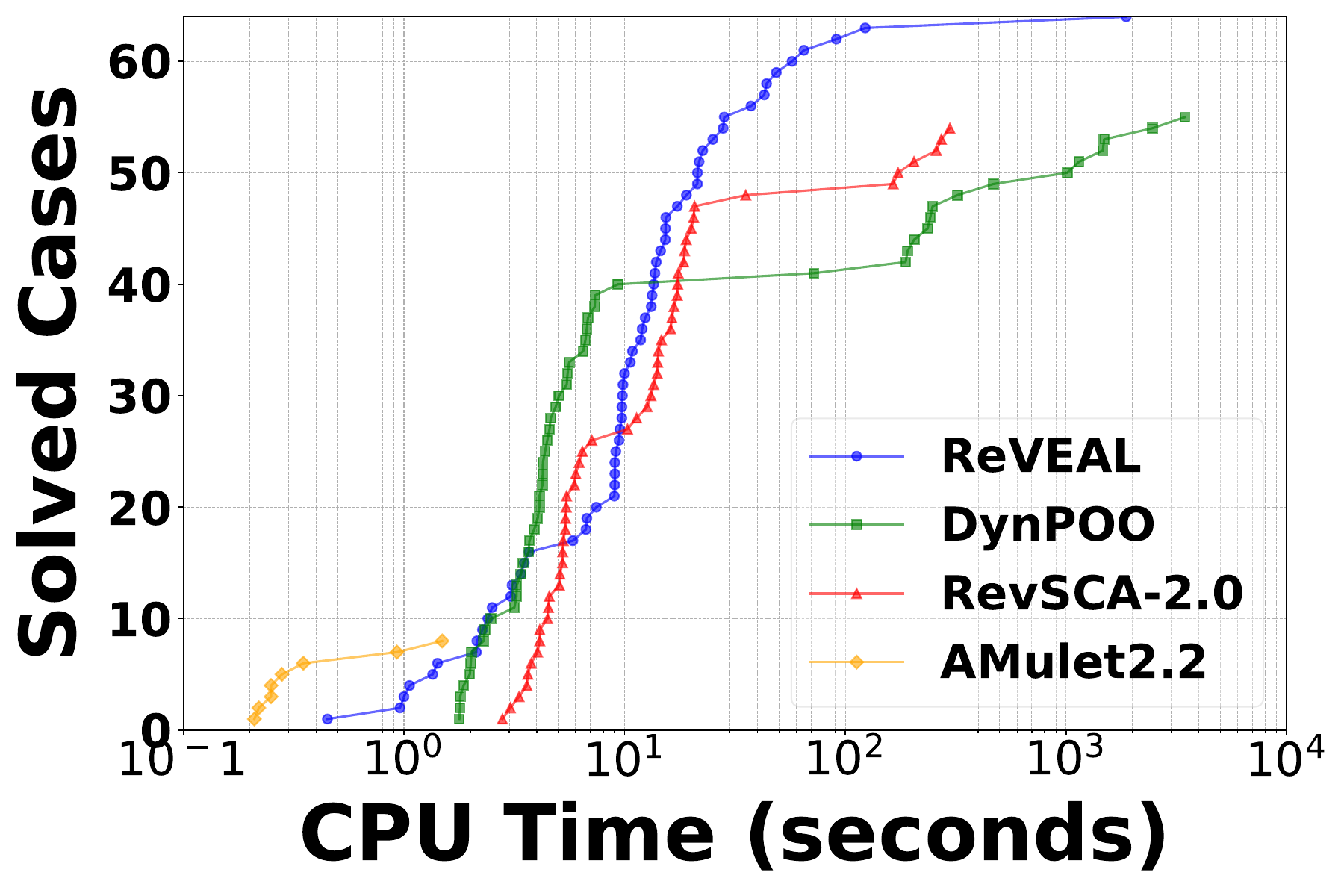}
        \caption{\mbox{resyn3\mbox{-}64}}
        \label{fig:first_image_time}
    \end{subfigure}
    \begin{subfigure}[b]{0.24\linewidth}
        \centering
        \includegraphics[width=\textwidth]{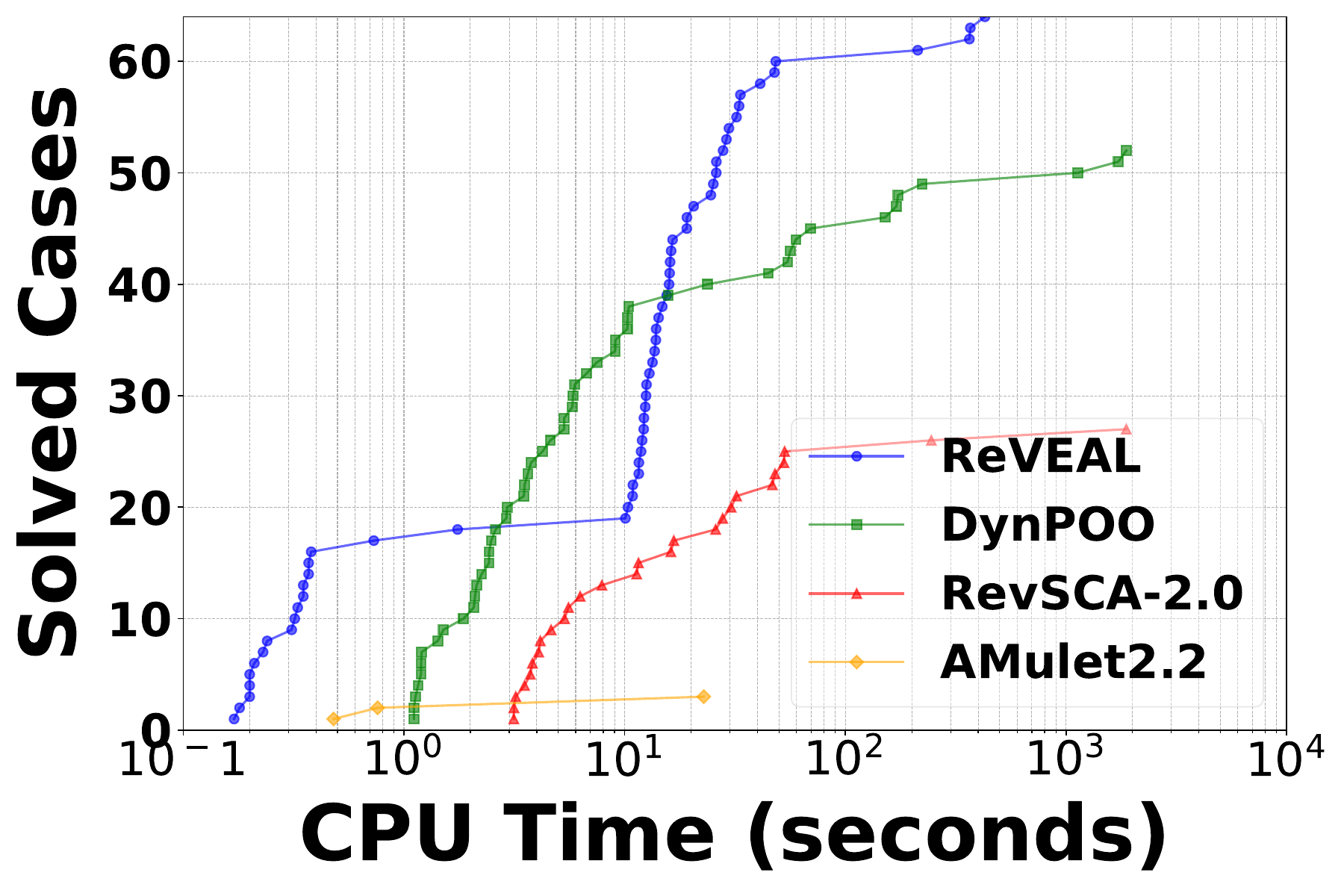}
        \caption{\mbox{dc2\mbox{-}64}}
        \label{fig:second_image_time}
    \end{subfigure}
    \begin{subfigure}[b]{0.24\linewidth}
        \centering
        \includegraphics[width=\textwidth]{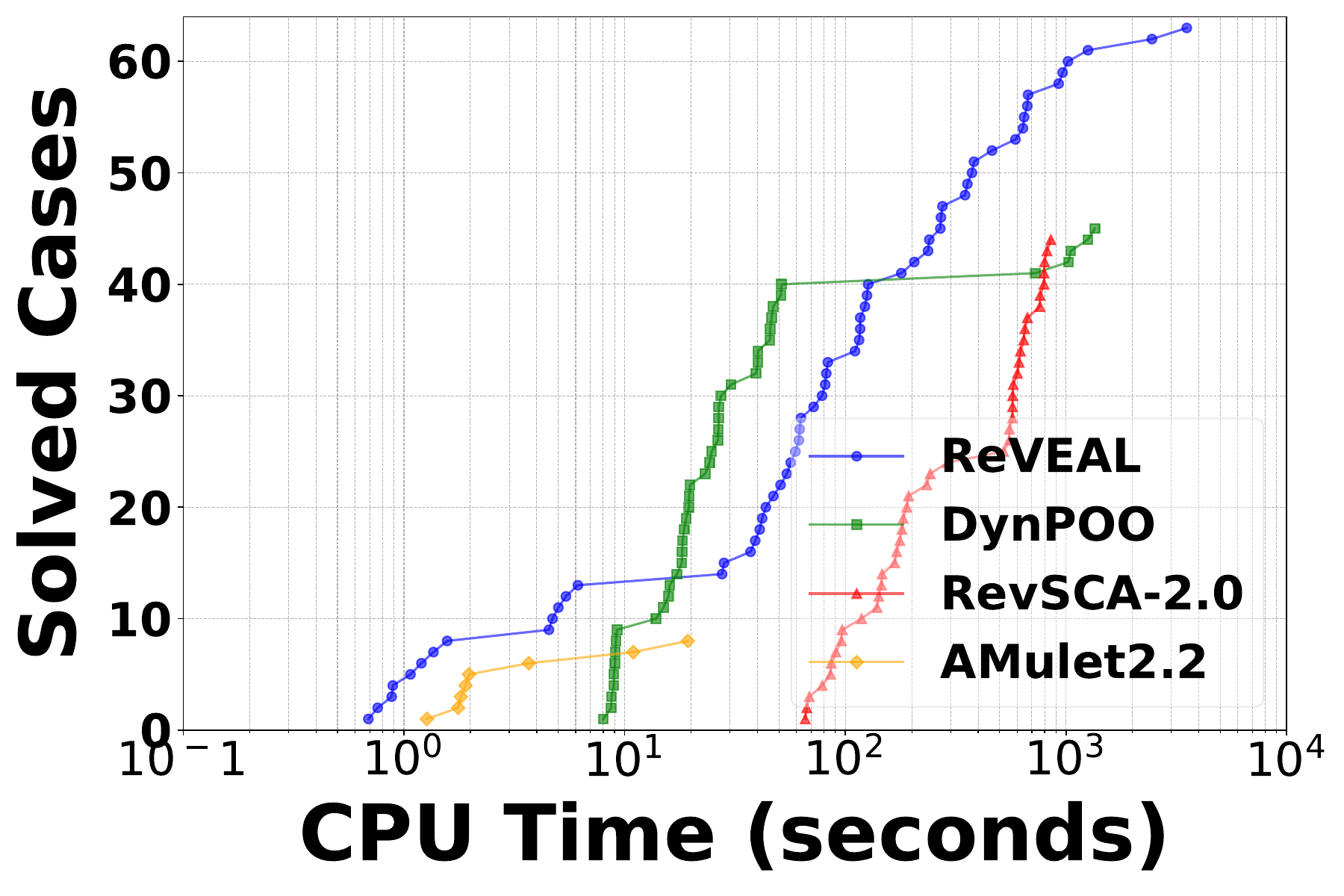}
        \caption{\mbox{resyn3\mbox{-}128}}
        \label{fig:third_image_time}
    \end{subfigure}
    \begin{subfigure}[b]{0.24\linewidth}
        \centering
        \includegraphics[width=\textwidth]{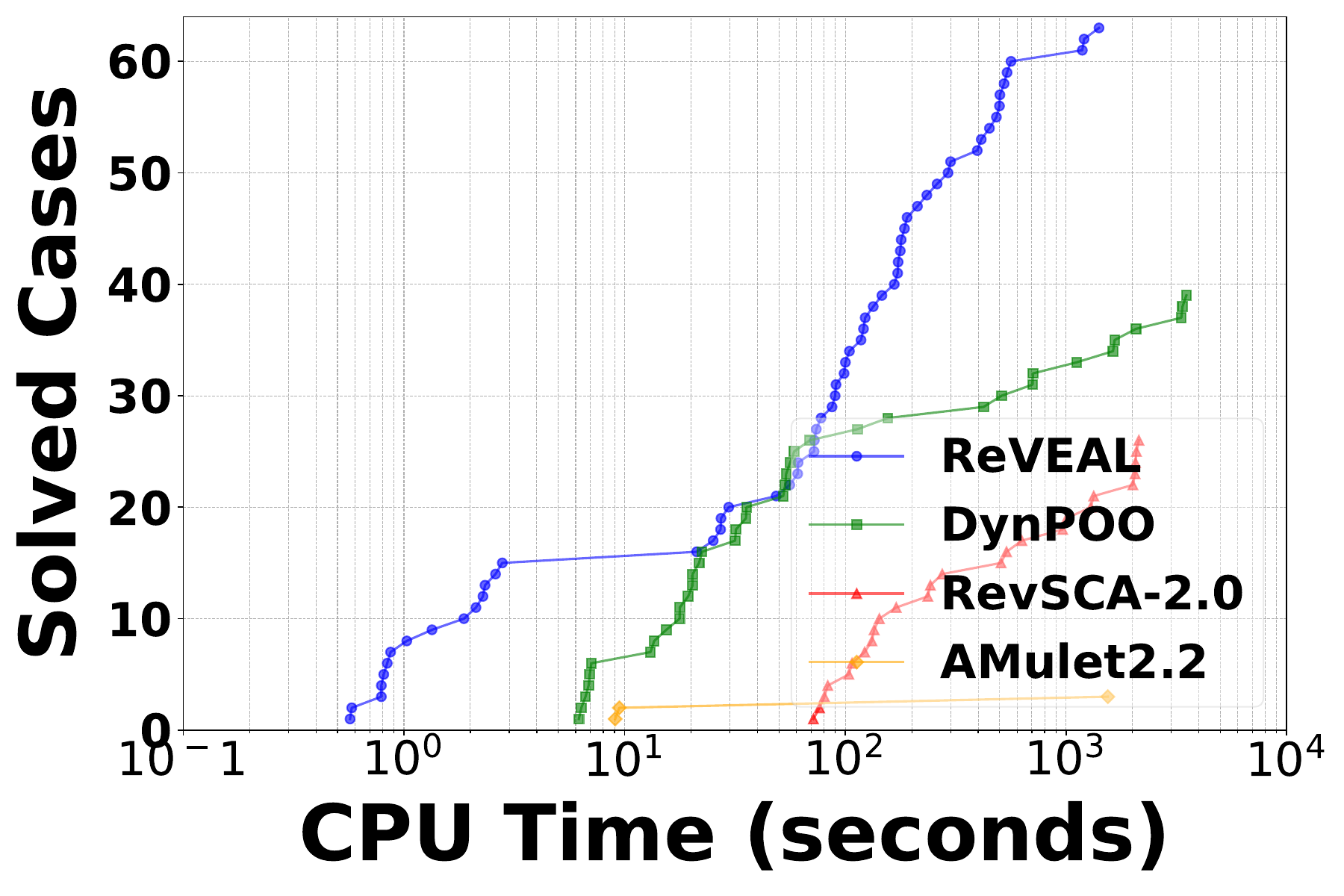}
        \caption{\mbox{dc2\mbox{-}128}}
        \label{fig:fourth_image_time}
    \end{subfigure}

    \begin{subfigure}[b]{0.24\linewidth}
        \centering
        \includegraphics[width=\textwidth]{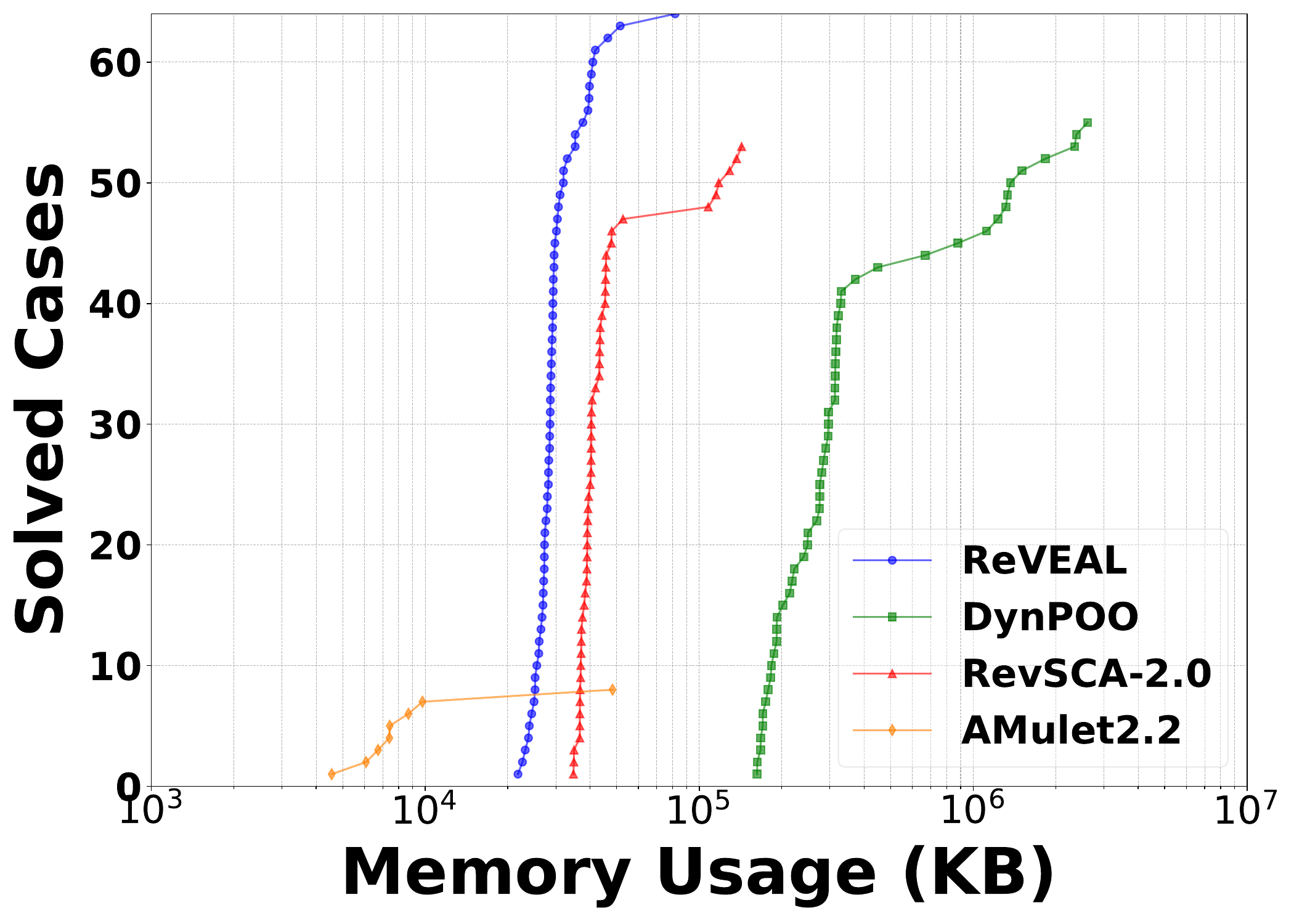}
        \caption{\mbox{resyn3\mbox{-}64}}
        \label{fig:first_image_memory}
    \end{subfigure}
    \hspace{-0.5em}
    \begin{subfigure}[b]{0.24\linewidth}
        \centering
        \includegraphics[width=\textwidth]{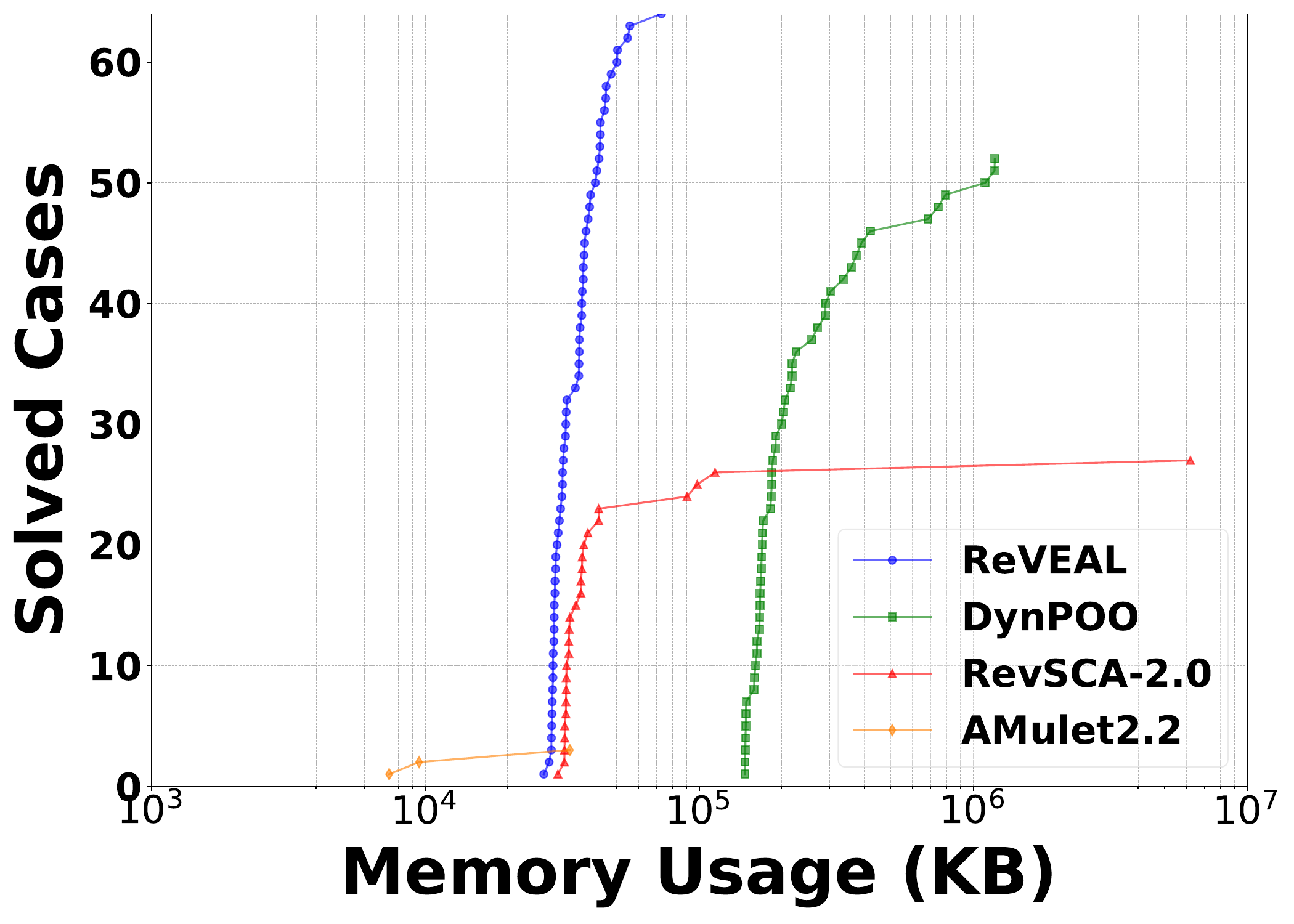}
        \caption{\mbox{dc2\mbox{-}64}}
        \label{fig:second_image_memory}
    \end{subfigure}
    \hspace{-0.5em}
    \begin{subfigure}[b]{0.24\linewidth}
        \centering
        \includegraphics[width=\textwidth]{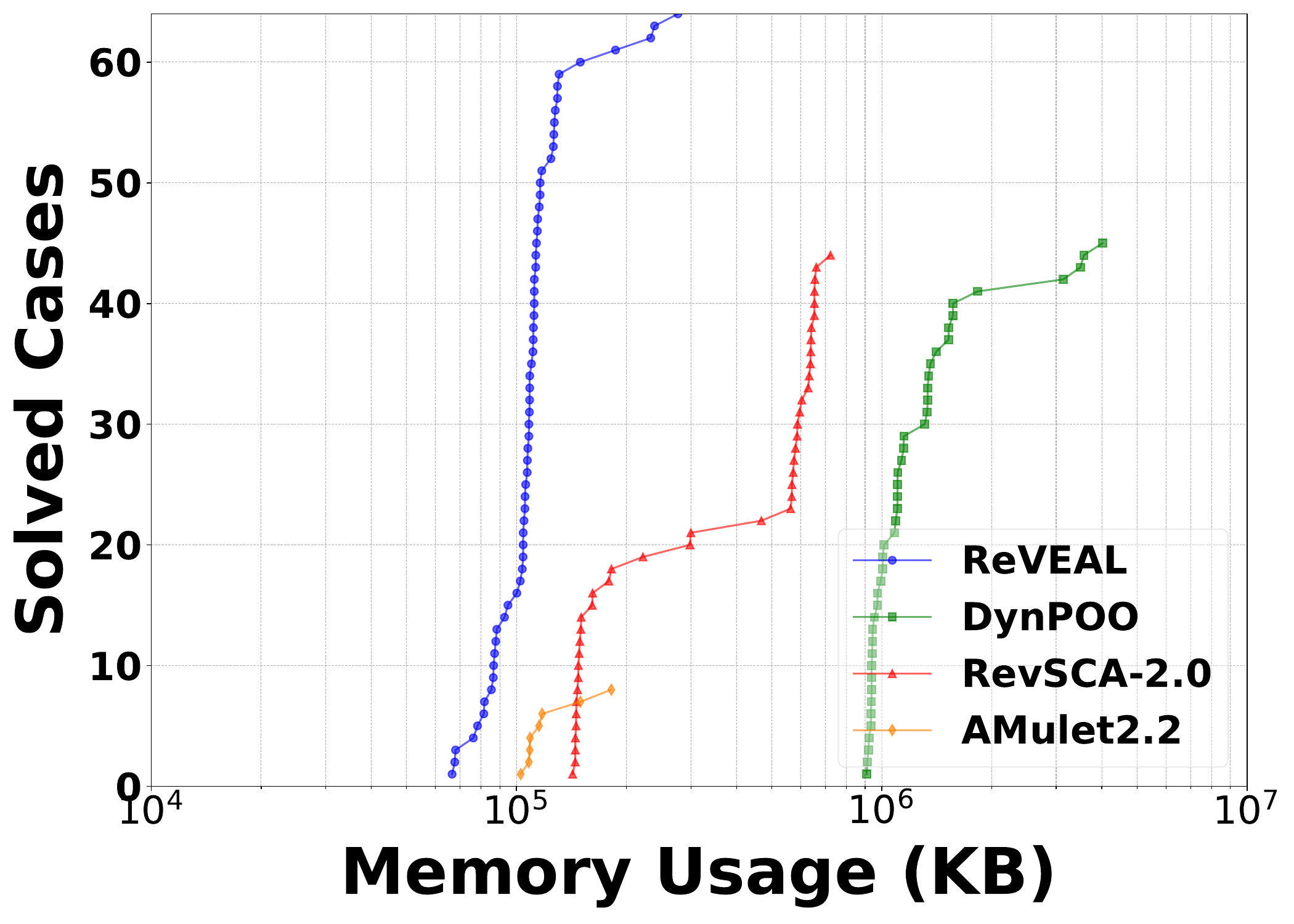}
        \caption{\mbox{resyn3\mbox{-}128}}
        \label{fig:third_image_memory}
    \end{subfigure}
    \hspace{-0.5em}
    \begin{subfigure}[b]{0.24\linewidth}
        \centering
        \includegraphics[width=\textwidth]{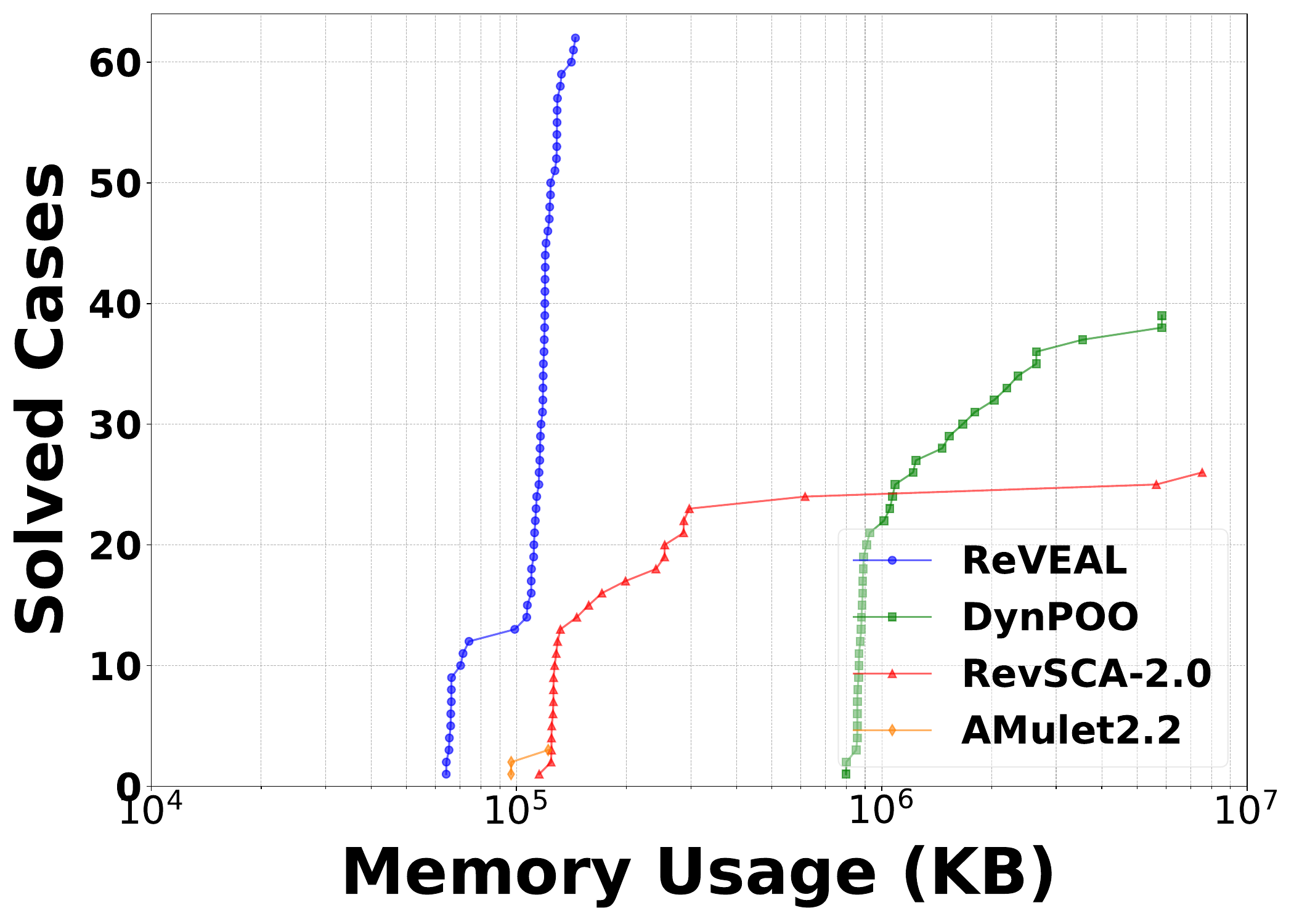}
        \caption{\mbox{dc2\mbox{-}128}}
        \label{fig:fourth_image_memory}
    \end{subfigure}

    \caption{Runtime and memory usage across different ``optimization-bitwidths''.}
    \vspace{-2ex}
    \label{fig:combined_images}

\end{figure*}
We systematically test all 64 possible template combinations and successfully verify equivalence between the top three predicted templates and the circuit under test. For a fair comparison as in~\cite{mahzoon2021revsca,konrad2024symbolic}, we use the same optimizations, \texttt{dc2} and \texttt{resyn3} in ABC. Experimental results in \Cref{tab:tool_results} and \Cref{fig:combined_images} show that ReVEAL successfully solves nearly all benchmarks, except one limited by excessive SAT-solving time. In general, minor errors in the FSA stage prediction are tolerable—SAT solving and sweeping remain highly effective even across structurally different FSA templates, providing a robustness cushion against FSA misclassifications.

\begin{table}[tb]
    \centering
    \caption{The number of solvable cases for different tools across benchmarks}
    \resizebox{0.6\textwidth}{!}{%
    \begin{tabular}{l||l|c|c|c|c}
        \hline\hline
        \multirow{3}{*}{\textbf{Tool}} & \multirow{3}{*}{\textbf{Result}} & \multicolumn{4}{c}{\textbf{Benchmark}} \\
        \cline{3-6}
        & & \multicolumn{2}{c|}{\textbf{64bit}} & \multicolumn{2}{c}{\textbf{128bit}} \\
         \cline{3-6}
        & & \textbf{dc2} & \textbf{resyn3} & \textbf{dc2} & \textbf{resyn3} \\
        \hline\hline

        \multirow{4}{*}{\textbf{AMulet-2.2}} 
        & Solved & 3/64 & 8/64 & 3/64 & 8/64 \\
        & TO & \color{red}61 & \color{red}55 & \color{red}61 & \color{red}56 \\
        & MO & \cellcolor{blue!15}0 & \color{red}1 & \cellcolor{blue!15}0 & \cellcolor{blue!15}0 \\
        & ``Circuit buggy'' & \cellcolor{blue!15}0 & \cellcolor{blue!15}0 & \cellcolor{blue!15}0 & \cellcolor{blue!15}0 \\
        \hline
        \multirow{4}{*}{\textbf{RevSCA-2.0}} 
        & Solved & 26/64 & 54/64 & 26/64 & 44/64 \\
        & TO & \color{red}7 & \color{red}3 & \color{red}7 & \color{red}8 \\
        & MO & \color{red}21 & \color{red}7 & \color{red}29 & \color{red}7 \\
        & ``Circuit buggy'' & \color{red}2 & \cellcolor{blue!15}0 & \color{red}2 & \color{red}5 \\
        \hline
        \multirow{4}{*}{\textbf{DynPOO}} 
        & Solved & 52/64 & 55/64 & 39/64 & 45/64 \\
        & TO & \color{red}12 & \color{red}9 & \color{red}22 & \color{red}7 \\
        & MO & \cellcolor{blue!15}0 & \cellcolor{blue!15}0 & \color{red}3 & \color{red}12 \\
        & ``Circuit buggy'' & \cellcolor{blue!15}0 & \cellcolor{blue!15}0 & \cellcolor{blue!15}0 & \cellcolor{blue!15}0 \\
        \hline
        \multirow{5}{*}{\textbf{ReVEAL}} 
        & Solved & \cellcolor{blue!15}\textbf{64/64} & \cellcolor{blue!15}\textbf{64/64} & \cellcolor{blue!15}\textbf{64/64} & \cellcolor{blue!15}\textbf{63/64} \\
        & TO & \cellcolor{blue!15}0 & \cellcolor{blue!15}0 & \cellcolor{blue!15}0 & \cellcolor{blue!15}\color{red}1 \\
        & MO & \cellcolor{blue!15}0 & \cellcolor{blue!15}0 & \cellcolor{blue!15}0 & \cellcolor{blue!15}0 \\
        & ``Circuit buggy'' & \cellcolor{blue!15}0 & \cellcolor{blue!15}0 & \cellcolor{blue!15}0 & \cellcolor{blue!15}0 \\
        & Stage FSA acc. (\%) & 96.88 & 100.00 & 92.19 & 100.00 \\
        \hline\hline
    \end{tabular}
    }
    \label{tab:tool_results}
    \vspace{-4ex}
\end{table}

\begin{table}[b]
 \vspace{-4ex}
  \centering
  \small
  \caption{Effect of missclassifications.}
  \setlength{\tabcolsep}{4pt}
  \renewcommand{\arraystretch}{1.2}

  \begin{minipage}[t]{0.48\linewidth}
    \centering
    \resizebox{0.8\textwidth}{!}{%
    \begin{tabular}{@{}r l l r@{}}
      \toprule
      \multicolumn{1}{c}{\#} & \multicolumn{1}{c}{Test circuit} & \multicolumn{1}{c}{Template} & \multicolumn{1}{c}{Runtime} \\
      \midrule
      1  & SP\_AR\_BK & SP\_AR\_BK & \cellcolor{blue!15}3.04 \\
      2  & SP\_AR\_BK & SP\_AR\_CK & 3.37 \\
      3  & SP\_AR\_BK & SP\_AR\_CL & 5083.42 \\
      4  & SP\_AR\_BK & SP\_AR\_KS & 232.18 \\
      5  & SP\_AR\_BK & SP\_AR\_LF & 33.46 \\
       6  & SP\_AR\_BK & SP\_AR\_RC & 3.37 \\
      7  & SP\_AR\_BK & SP\_AR\_SE & 3.15 \\
      \bottomrule
    \end{tabular}}
    \caption*{(a) Missclassified FSA.}\label{tbl:missa}
  \end{minipage}%
  \hfill
  \begin{minipage}[t]{0.48\linewidth}
    \centering
    \resizebox{0.8\textwidth}{!}{%
    \begin{tabular}{@{}r l l r@{}}
      \toprule
      \multicolumn{1}{c}{\#} & \multicolumn{1}{c}{Test circuit} & \multicolumn{1}{c}{Template} & \multicolumn{1}{c}{Runtime} \\
      \midrule
      1  & SP\_DT\_BK & SP\_DT\_BK &  \cellcolor{blue!15}3.8 \\
      2  & SP\_DT\_BK & SP\_AR\_BK & \textcolor{red}{TO} \\
      3  & SP\_DT\_BK & SP\_CWT\_BK & \textcolor{red}{TO} \\
      4 & SP\_DT\_BK & SP\_WT\_BK & \textcolor{red}{TO} \\
      \\
      \\
      \\
      \bottomrule
    \end{tabular}}
    \caption*{(b) Missclassified PPA.}\label{tbl:missb}
  \end{minipage}
  \label{tbl:mispredict}
  \vspace{-2ex}
\end{table}

Specifically, we highlight the comparison of verification times and memory usage across various tools in~\Cref{fig:combined_images}. We include per-case solving details in \Cref{tab:test_resyn3_appendix}.
In general, existing CA tools suffer from vanishing monomial computations, which make them inefficient for handling large optimized circuits.
On average, our method runs in 174.34 seconds using 75.01 MB memory, while \dynphaseorderopt~\cite{konrad2024symbolic} requires 936.40 seconds and 1,374.40 MB. This demonstrates a 5.37× speed improvement and 18.33× memory reduction in favor of our approach.
As can be seen in~\Cref{tab:tool_results}, the tool \revscatwo reports correct circuits as ``buggy''. 
This is because \revscatwo has bugs in reversely identifying functional arithmetic blocks, which cause the circuit to become inequivalent to the original circuit.
 \vspace{-2ex}
\paragraph{Impact of Misidentification on Verification Runtime.}
As an illustration, we run the following experiment in Table~\ref{tbl:mispredict}. 
We fix a 64-bit test multiplier that is optimized with resyn3 (column ``Test circuit'') and use equivalence checking to compare it against multipliers of different architectures. In Table~\ref{tbl:missa}(a), we resemble the situation when FSA (stage 3) is mispredicted, and Table~\ref{tbl:missa}(b) mimics the case when PPA (stage 2) is mispredicted. Note that not all FSA misprediction results in long runtime (see case 6 and 7). 
Despite that PPA misprediction leads to excessively long solving time, 
as shown by the earlier experiment in Table~\ref{tab:stage_accuracy}, 
our model achieves a high accuracy in PPG and PPA prediction and this helps avoid the long equivalence checking time.
 \vspace{-2ex}
\paragraph{Buggy-circuit robustness.}
We also evaluate \textsc{ReVEAL} on buggy designs. For each optimized 64-bit AIG, we
create buggy variants by rewiring the fanins of three internal nodes (preserving
acyclicity) and resample until random simulation finds a mismatch at some primary
output. We then run SAT-based CEC against the corresponding template. ABC's
\texttt{cec} takes 0.05\,s on average (max 1.98\,s), and all instances are \texttt{SAT}
(non-equivalence detected).

\paragraph{Certification.}
Once \textsc{ReVEAL} concludes equivalence, we can provide proof certificates for both
verification steps: (i) each unoptimized template in our library is certified by
\amulettwo via an algebraic proof certificate (a sequence of polynomial additions and
multiplications)~\cite{kaufmannPracticalAlgebraicCalculus2022}; and (ii) SAT-based CEC
using \kissat outputs a DRUP clausal proof certificate. Both certificate types can be
checked by certified proof checkers~\cite{kaufmannPracticalAlgebraicCalculus2022,tan-cakeml23}.
\vspace{-2ex}



\paragraph{Comparison with CA+SAT Methods on Extensively Optimized Multipliers.}
\label{Q3_2}
In~\Cref{tab:RefSCAT-reveal}, we further evaluate extensively optimized\footnote{The multipliers are optimized using the following sequence in ABC, applied twice: {\scriptsize \texttt{\&mfs: logic; mfs2 -W 20; mfs; st; dc2 -l; resub -l -K 16 -N 3 -w 100; logic; mfs2 -W 20; mfs; st; resyn -l; resyn; resyn2; resyn3; dc2 -l}}} multipliers, comparing ReVEAL with RefSCAT~\cite{li2024refscat}, a method that combines SAT and CA by reconstructing complete references through adder boundary detection using a constraint satisfaction algorithm. For training and testing, we use the same settings as above in \Cref{exp_setting}. Training uses even bit-width, ranging from 32 to 60 bits and testing uses unseen 64, 128, and 256-bit designs. Since RefSCAT is not publicly available, we use the experimental data from~\cite{li2024refscat} for comparison. 
In terms of runtime, ReVEAL outperforms RefSCAT by a factor of 13.39, despite one corner case slowed down by SAT sweeping time. 
The results show that ReVEAL’s time overhead is almost entirely spent on equivalence checking, while the preprocessing phase takes less than 1 second. In contrast, RefSCAT uses significant time on adder boundary detection, particularly in highly optimized cases. Notably, RefSCAT currently cannot verify Booth-encoded or non-adder-based multipliers, requiring online RTL multiplier reconstruction that introduces RTL to AIG conversion overhead and online computer-algebraic verification runtime. Our approach, on the other hand, is faster and more versatile for reverse engineering and architectural inference. 
\begin{table*}[tb]
\caption{Time Comparison between RefSCAT and ReVEAL.}
\centering
\resizebox{1\linewidth}{!}{%
\begin{tabular}{l||r|rrrrr}
\hline \hline
\multicolumn{1}{c||}{\multirow{2}{*}{Benchmark}} &
\multicolumn{1}{c|}{\textbf{RefSCAT}} &
\multicolumn{5}{c}{\textbf{ReVEAL}} \\
\cline{2-7}
& \multicolumn{1}{c|}{Time (s)} &
\multicolumn{1}{c}{CE (s)} &
\multicolumn{1}{c}{FE (s)} &
\multicolumn{1}{c}{Inference (s)} &
\multicolumn{1}{c}{EQ (s)} &
\multicolumn{1}{c}{Total Time (s)} \\
\hline \hline
128\_128\_U\_SP\_DT\_HCA & 2518.8 & 0.51 & 0.10 & 0.01 & 40.13 & \cellcolor{blue!15}\textbf{40.74} \\ \hline
128\_128\_U\_SP\_WT\_HCA & 1791.1 & 0.49 & 0.11 & 0.01 & 42.69 & \cellcolor{blue!15}\textbf{43.29} \\ \hline
128\_128\_U\_SP\_DT\_CK & 144.1 & 0.49 & 0.10 & 0.01 & 24.84 & \cellcolor{blue!15}\textbf{25.44} \\ \hline
128\_128\_U\_SP\_DT\_KS & 156.4 & 0.56 & 0.08 & 0.01 & 29.00 & \cellcolor{blue!15}\textbf{29.65} \\ \hline
128\_128\_U\_SP\_WT\_CK & 403.2 & 0.54 & 0.10 & 0.01 & 108.51 & \cellcolor{blue!15}\textbf{109.16} \\ \hline
128\_128\_U\_SP\_WT\_KS & 422.1 & 0.51 & 0.09 & 0.01 & 54.11 & \cellcolor{blue!15}\textbf{54.71} \\ \hline
128\_128\_U\_SP\_WT\_LF & 1889.2 & 0.49 & 0.11 & 0.01 & 41.21 & \cellcolor{blue!15}\textbf{41.83} \\ \hline
128\_128\_U\_SP\_DT\_CL & 127.2 & 0.50 & 0.08 & 0.01 & 211.07 & 211.66 \\ \hline\hline
\end{tabular}}
\label{tab:RefSCAT-reveal}
\textbf{\textsuperscript{*}} CE: Cone Extraction; FE: Feature Extraction; EQ: Equivalence Checking;
\vspace{-2ex}
\end{table*}

\qst{Are the proposed cone extraction, feature extraction, and neural network models domain-specifically helpful in the task?}\label{exp:RQ3}

To evaluate the contribution of each technique in ReVEAL, we conduct an ablation study by selectively removing key components of our approach. The results are shown in~\Cref{tab:performance_comparison}. 
First, in \emph{ReVEAL w/o cone extraction}, we train the model on the entire large graph without applying cone extraction. This leads to a significant performance decline during both \emph{Stage PPA} and \emph{Stage FSA} inference, primarily due to the over-squeezing and over-smoothing problems that GNNs face when operating on large graphs, demonstrating the necessity of cone extraction for efficient processing. Second, in \emph{ReVEAL w/o word-level features}, we replace our word-level features with the basic functional AIG features used in \hoga and \gamora. This results in a decrease in accuracy, indicating that word-level feature extraction contributes positively to the model's performance. Last, we conduct a comprehensive comparison of our approach with existing mainstream \emph{GNN models}. Our analysis reveals that models lacking hierarchical classification exhibited higher misclassification rates for both non-tree and tree adders. In contrast, ReVEAL, which integrates hierarchical classification with bidirectional GraphSAGE, significantly reduces misclassification and enhances overall classification accuracy. 
\begin{table}[tb]
\centering
\caption{Ablation Study and Performance Comparison of Different Methods.\label{tab:performance_comparison}}
\resizebox{0.8\textwidth}{!}{%
\begin{threeparttable}
\begin{tabular}{l||l|c|c|c|c}
\hline\hline
\multirow{2}{*} {}& \multirow{2}{*}{\textbf{Method}} & \multirow{2}{*}{\textbf{Stage}} & \multicolumn{3}{c}{\textbf{Accuracy (\%)}} \\
& & & \textbf{TOP 1} & \textbf{TOP 2} & \textbf{TOP 3} \\
\hline\hline
\multirow{2}{*}{\textbf{Base}}
& \cellcolor{blue!15}\textbf{ReVEAL} & Stage PPA & \cellcolor{blue!15}\textbf{1.0} & / & / \\
\cline{2-6}
& \cellcolor{blue!15}\textbf{ReVEAL} & Stage FSA & \cellcolor{blue!15}\textbf{0.85} & \cellcolor{blue!15}\textbf{0.97} & \cellcolor{blue!15}\textbf{0.98} \\
\hline\hline
\multirow{2}{*}{\textbf{w/o CE\textsuperscript{*}}}
& ReVEAL (w/o CE\textsuperscript{*}) & Stage PPA & 0.15 & / & / \\
\cline{2-6}
& ReVEAL (w/o CE\textsuperscript{*}) & Stage FSA & 0.16 & 0.28 & 0.40 \\
\hline\hline
\multirow{2}{*}{\textbf{w/o WF\textsuperscript{*}}}
& ReVEAL (w/o WF\textsuperscript{*}) & Stage PPA & \cellcolor{blue!15}\textbf{1.0} & / & / \\
\cline{2-6}
& ReVEAL (w/o WF\textsuperscript{*}) & Stage FSA & 0.76 & 0.94 & 0.95 \\
\hline\hline
\multirow{6}{*}{\textbf{GNN Model}}
& GCN & \multirow{6}{*}{Stage FSA} & 0.72 & 0.9 & 0.94 \\
& GAT &  & 0.75 & 0.91 & 0.95 \\
& GraphSAGE &  & 0.80 & 0.96 & 0.97 \\
& SagePooling~\cite{DBLP:conf/nips/KnyazevTA19} &  & 0.71 & 0.91 & 0.92 \\
& HOGA~\cite{deng2024less} &  & 0.79 & 0.92 & 0.94 \\
& Graph Transformer~\cite{DBLP:conf/ijcai/ShiHFZWS21} &  & 0.64 & 0.86 & 0.91 \\
\hline\hline
\end{tabular}
\begin{tablenotes}
\small
\item[$^{*}$]CE: Cone Extraction; WF: Word Feature Extraction
\end{tablenotes}
\end{threeparttable}
}
\vspace{-2ex}
\end{table}






\section{Conclusion and Future Work}
\label{sec:conclusion}
In this paper, we have introduced ReVEAL, the first GNN-based framework for architectural reasoning in the reverse engineering of optimized multipliers. ReVEAL integrates critical cone extraction, word-level feature encoding, and a hierarchical multi-task classification model to achieve strong generalizability and high accuracy. Training scales well when bit-widths grow because critical cone extraction fixes the learning scope to small, stage-specific subgraphs.
By combining GNN-based reverse engineering with modern SAT solvers and computer algebra verifiers, ReVEAL effectively overcomes the performance bottlenecks inherent in traditional CA and CA+SAT-based verification methods for optimized multiplier verification.
A future extension for the work includes applying the critical-cone learning paradigm to MAC units and other more complex datapath circuits.\looseness=-1


\bibliography{refs,yu-add}
\newpage
\appendix
\section{Word-Level Arithmetic Block Identification via Cut Enumeration}

To enable word-level feature encoding on an optimized AIG, we first perform reverse
identification of small arithmetic primitives (XOR/MAJ, half adders, and full adders)
using \emph{cut enumeration}. We briefly recall the notions of $k$-feasible cuts and
cut enumeration in the AIG setting. For readers interested in further algorithmic details, we refer to
Section~3 (\emph{Cut Computation}) of~\cite{chatterjee2007algorithms}; moreover, the
ABC codebase provides additional practical examples of arithmetic reverse
engineering built on cut enumeration (see the \texttt{acec} directory).

\paragraph{\textbf{Cuts and $k$-feasible cuts in an AIG.}}
Let $G$ be an AIG and let $v$ be a node in $G$. A \emph{cut} $c$ of $v$ is a set of nodes
in the transitive fan-in of $v$ such that every path from any primary input to $v$
passes through at least one node in $c$. A cut is \emph{irredundant} if none of its proper
subsets is also a cut of $v$. A cut $c$ is a \emph{$k$-feasible cut} if it is irredundant and
$|c|\le k$. \emph{Cut enumeration} refers to enumerating all $k$-feasible cuts for each node,
which provides a compact way to represent candidate local subcircuits rooted at $v$.

\paragraph{\textbf{Truth-table extraction and Boolean-function matching.}}
For each node $v$, we enumerate its $k$-feasible cuts and compute the truth table of
the corresponding cut function. We maintain an
offline precomputed database that stores the set of truth tables that are
\emph{NPN-equivalent} to target Boolean primitives, in particular XOR and MAJ
functions (e.g., XOR2/XOR3 and MAJ2/MAJ3), where NPN-equivalence accounts for input
negations, input permutations, and optional output negation. A cut is classified as an
\textsf{XOR} (resp. \textsf{MAJ}) candidate if its truth table matches any entry in the
corresponding NPN-class database.

\paragraph{\textbf{Recovering half adders and full adders.}}
After collecting all \textsf{XOR} and \textsf{MAJ} candidates, we identify half adders (HA)
and full adders (FA) by pairing compatible candidates that share the same input nodes.
Concretely, an HA is detected by a pair consisting of an \textsf{XOR2} node (sum) and an
\textsf{MAJ2}-equivalent node (carry) that share the same two inputs. An FA is
detected by a pair consisting of an \textsf{XOR3} node (sum) and an \textsf{MAJ3} node (carry)
that share the same three inputs. For each proposed (sum, carry) pair, we additionally
validate that, after an appropriate NPN transformation, the pair implements the
canonical HA/FA truth-table specification. Nodes covered by validated pairs are
annotated as HA/FA blocks; remaining matched nodes are annotated as standalone
\textsf{XOR} (or \textsf{MAJ}) primitives and used as word-level features in the subsequent
GNN inference.
\label{sec:word_level_id}
\section{Details of GNN Predictions}
In this appendix, we would like to provide the details of GNN predictions given all available templates.
The architecture is denoted by the following format ``FirstStage\_SecondStage\_LastStage'' representing the choices for PPG, PPA and FSA, respectively.
The multipliers are 64-bit wide optimized with dc2. As can be seen from Table~\ref{tab:dc2-64-gnn-rank-part1}, there is little challenge to predict PPG and PPA, while occasionally, FSA may be mispredicted.

\begin{table}[H]
  \centering
  \small
  \caption{Top-3 predictions for dc2-optimized circuits}
  \label{tab:dc2-64-gnn-rank-part1}
  \setlength{\tabcolsep}{4pt}
  \renewcommand{\arraystretch}{1.2}
  \begin{tabular}{@{}r l l l l@{}}
    \toprule
    \multicolumn{1}{c}{\#} & \multicolumn{1}{c}{Tested dc2-opt circuit} & \multicolumn{1}{c}{rank-1 prediction} & \multicolumn{1}{c}{rank-2 prediction} & \multicolumn{1}{c}{rank-3 prediction} \\
    \midrule
    1  & SP\_4to2\_BK & SP\_4to2\_BK & SP\_4to2\_LF & SP\_4to2\_JCA \\
    2  & SP\_4to2\_HCA & SP\_4to2\_HCA & SP\_4to2\_BK & SP\_4to2\_LF \\
    3  & SP\_4to2\_JCA & SP\_4to2\_LF & SP\_4to2\_JCA & SP\_4to2\_KS \\
    4  & SP\_4to2\_KS & SP\_4to2\_KS & SP\_4to2\_HCA & SP\_4to2\_LF \\
    5  & SP\_4to2\_LF & SP\_4to2\_LF & SP\_4to2\_JCA & SP\_4to2\_KS \\
    6  & SP\_4to2\_RC & SP\_4to2\_RC & SP\_4to2\_CK & SP\_4to2\_CL \\
    7  & SP\_DT\_BK & SP\_DT\_BK & SP\_DT\_LF & SP\_DT\_JCA \\
    8  & SP\_DT\_HCA & SP\_DT\_HCA & SP\_DT\_KS & SP\_DT\_LF \\
    9  & SP\_DT\_JCA & SP\_DT\_LF & SP\_DT\_JCA & SP\_DT\_HCA \\
    10 & SP\_DT\_KS & SP\_DT\_KS & SP\_DT\_HCA & SP\_DT\_LF \\
    11 & SP\_DT\_LF & SP\_DT\_LF & SP\_DT\_JCA & SP\_DT\_HCA \\
    12 & SP\_DT\_RC & SP\_DT\_RC & SP\_DT\_CK & SP\_DT\_CL \\
    13 & SP\_WT\_BK & SP\_WT\_BK & SP\_WT\_LF & SP\_WT\_JCA \\
    14 & SP\_WT\_HCA & SP\_WT\_HCA & SP\_WT\_BK & SP\_WT\_LF \\
    15 & SP\_WT\_JCA & SP\_WT\_LF & SP\_WT\_JCA & SP\_WT\_BK \\
    16 & SP\_WT\_KS & SP\_WT\_KS & SP\_WT\_HCA & SP\_WT\_RC \\
    17 & SP\_WT\_LF & SP\_WT\_LF & SP\_WT\_JCA & SP\_WT\_BK \\
    18 & SP\_WT\_RC & SP\_WT\_RC & SP\_WT\_CK & SP\_WT\_CL \\
    19 & BP\_4to2\_BK & BP\_4to2\_BK & BP\_4to2\_LF & BP\_4to2\_JCA \\
    20 & BP\_4to2\_HCA & BP\_4to2\_HCA & BP\_4to2\_KS & BP\_4to2\_BK \\
    21 & BP\_4to2\_JCA & BP\_4to2\_LF & BP\_4to2\_JCA & BP\_4to2\_KS \\
    22 & BP\_4to2\_KS & BP\_4to2\_KS & BP\_4to2\_HCA & BP\_4to2\_LF \\
    23 & BP\_4to2\_LF & BP\_4to2\_LF & BP\_4to2\_JCA & BP\_4to2\_KS \\
    24 & BP\_4to2\_RC & BP\_4to2\_RC & BP\_4to2\_CK & BP\_4to2\_CL \\
    25 & BP\_DT\_BK & BP\_DT\_BK & BP\_DT\_LF & BP\_DT\_JCA \\
    26 & BP\_DT\_HCA & BP\_DT\_HCA & BP\_DT\_BK & BP\_DT\_LF \\
    27 & BP\_DT\_JCA & BP\_DT\_LF & BP\_DT\_JCA & BP\_DT\_KS \\
    28 & BP\_DT\_KS & BP\_DT\_KS & BP\_DT\_HCA & BP\_DT\_LF \\
    \bottomrule
  \end{tabular}
\end{table}

\begin{table}[H]
  \centering
  \small
  \setlength{\tabcolsep}{4pt}
  \renewcommand{\arraystretch}{1.2}
  \begin{tabular}{@{}r l l l l@{}}
    \toprule
    \multicolumn{1}{c}{\#} & \multicolumn{1}{c}{Tested dc2-opt circuit} & \multicolumn{1}{c}{rank-1 prediction} & \multicolumn{1}{c}{rank-2 prediction} & \multicolumn{1}{c}{rank-3 prediction} \\
    \midrule
    29 & BP\_DT\_LF & BP\_DT\_LF & BP\_DT\_JCA & BP\_DT\_KS \\
    30 & BP\_DT\_RC & BP\_DT\_RC & BP\_DT\_CK & BP\_DT\_CL \\
    31 & BP\_WT\_BK & BP\_WT\_BK & BP\_WT\_LF & BP\_WT\_HCA \\
    32 & BP\_WT\_HCA & BP\_WT\_HCA & BP\_WT\_LF & BP\_WT\_BK \\
    33 & BP\_WT\_JCA & BP\_WT\_LF & BP\_WT\_JCA & BP\_WT\_BK \\
    34 & BP\_WT\_KS & BP\_WT\_KS & BP\_WT\_HCA & BP\_WT\_RC \\
    35 & BP\_WT\_LF & BP\_WT\_LF & BP\_WT\_JCA & BP\_WT\_BK \\
    36 & BP\_WT\_RC & BP\_WT\_RC & BP\_WT\_CK & BP\_WT\_CL \\
    37 & SP\_AR\_BK & SP\_AR\_SE & SP\_AR\_CL & SP\_AR\_HCA \\
    38 & SP\_AR\_CK & SP\_AR\_CK & SP\_AR\_RC & SP\_AR\_CL \\
    39 & SP\_AR\_CL & SP\_AR\_SE & SP\_AR\_CL & SP\_AR\_HCA \\
    40 & SP\_AR\_KS & SP\_AR\_KS & SP\_AR\_LF & SP\_AR\_RC \\
    41 & SP\_AR\_LF & SP\_AR\_SE & SP\_AR\_CL & SP\_AR\_HCA \\
    42 & SP\_AR\_RC & SP\_AR\_RC & SP\_AR\_CL & SP\_AR\_CK \\
    43 & SP\_AR\_SE & SP\_AR\_SE & SP\_AR\_CL & SP\_AR\_HCA \\
    44 & SP\_CWT\_BK & SP\_CWT\_BK & SP\_CWT\_LF & SP\_CWT\_JCA \\
    45 & SP\_CWT\_CK & SP\_CWT\_RC & SP\_CWT\_CK & SP\_CWT\_CL \\
    46 & SP\_CWT\_CL & SP\_CWT\_CL & SP\_CWT\_SE & SP\_CWT\_HCA \\
    47 & SP\_CWT\_KS & SP\_CWT\_KS & SP\_CWT\_HCA & SP\_CWT\_LF \\
    48 & SP\_CWT\_LF & SP\_CWT\_LF & SP\_CWT\_JCA & SP\_CWT\_BK \\
    49 & SP\_CWT\_RC & SP\_CWT\_RC & SP\_CWT\_CK & SP\_CWT\_CL \\
    50 & SP\_CWT\_SE & SP\_CWT\_SE & SP\_CWT\_CL & SP\_CWT\_HCA \\
    51 & SP\_DT\_BK & SP\_DT\_BK & SP\_DT\_SE & SP\_DT\_RC \\
    52 & SP\_DT\_CK & SP\_DT\_CK & SP\_DT\_RC & SP\_DT\_CL \\
    53 & SP\_DT\_CL & SP\_DT\_CL & SP\_DT\_LF & SP\_DT\_HCA \\
    54 & SP\_DT\_KS & SP\_DT\_KS & SP\_DT\_HCA & SP\_DT\_RC \\
    55 & SP\_DT\_LF & SP\_DT\_KS & SP\_DT\_LF & SP\_DT\_JCA \\
    56 & SP\_DT\_RC & SP\_DT\_RC & SP\_DT\_CK & SP\_DT\_CL \\
    57 & SP\_DT\_SE & SP\_DT\_SE & SP\_DT\_CL & SP\_DT\_HCA \\
    58 & SP\_WT\_BK & SP\_WT\_BK & SP\_WT\_LF & SP\_WT\_JCA \\
    59 & SP\_WT\_CK & SP\_WT\_CK & SP\_WT\_RC & SP\_WT\_CL \\
    60 & SP\_WT\_CL & SP\_WT\_CL & SP\_WT\_LF & SP\_WT\_HCA \\
    61 & SP\_WT\_KS & SP\_WT\_KS & SP\_WT\_HCA & SP\_WT\_RC \\
    62 & SP\_WT\_LF & SP\_WT\_LF & SP\_WT\_JCA & SP\_WT\_HCA \\
    63 & SP\_WT\_RC & SP\_WT\_RC & SP\_WT\_CK & SP\_WT\_CL \\
    64 & SP\_WT\_SE & SP\_WT\_SE & SP\_WT\_CL & SP\_WT\_HCA \\
    \bottomrule
  \end{tabular}

  \label{tab:dc2-64-gnn-rank-part2}
\end{table}
\section{Detailed Equivalence Checking Time}

This section presents a complete list of equivalence checking time, using 128-bit multipliers  optimized by  resyn3.


\begin{table}[H]
  \centering
  \caption{Time and memory consumption of equivalence checking}
  \small
  \setlength{\tabcolsep}{4pt}
  \renewcommand{\arraystretch}{1.2}
  \begin{tabular}{@{}r l l l l@{}}
    \toprule
    \multicolumn{1}{c}{\#} & \multicolumn{1}{c}{Tested\_resyn3\_128\_circuit} & \multicolumn{1}{c}{Time (seconds)} & \multicolumn{1}{c}{Memory (KB)} & \multicolumn{1}{c}{Method} \\
    \midrule
    1 & SP\_4to2\_BK & 61.60 & 78184 & kissat \\
    2 & SP\_4to2\_HCA & 81.04 & 87080 & sat-sweeping \\
    3 & SP\_4to2\_JCA & 125.33 & 81688 & kissat \\
    4 & SP\_4to2\_KS & 275.53 & 86572 & kissat \\
    5 & SP\_4to2\_LF & 126.82 & 86368 & kissat \\
    6 & SP\_4to2\_RC & 0.69 & 85356 & fraig \\
    7 & SP\_DT\_BK & 47.19 & 104264 & kissat \\
    8 & SP\_DT\_HCA & 81.94 & 105936 & kissat \\
    9 & SP\_DT\_JCA & 54.29 & 104828 & kissat \\
    10 & SP\_DT\_KS & 116.78 & 112904 & kissat \\
    11 & SP\_DT\_LF & 56.57 & 103688 & kissat \\
    12 & SP\_DT\_RC & 0.89 & 111632 & fraig \\
    13 & SP\_WT\_BK & 50.87 & 104216 & kissat \\
    14 & SP\_WT\_HCA & 5.42 & 67748 & sat-sweeping \\
    15 & SP\_WT\_JCA & 62.08 & 105424 & kissat \\
    16 & SP\_WT\_KS & 179.35 & 111812 & kissat \\
    17 & SP\_WT\_LF & 62.93 & 104300 & kissat \\
    18 & SP\_WT\_RC & 41.98 & 108100 & kissat \\
    19 & BP\_4to2\_BK & 83.30 & 129412 & sat-sweeping \\
    20 & BP\_4to2\_HCA & 271.10 & 111212 & sat-sweeping \\
    21 & BP\_4to2\_JCA & 925.40 & 100176 & kissat \\
    22 & BP\_4to2\_KS & 589.88 & 126496 & sat-sweeping \\
    23 & BP\_4to2\_LF & 964.83 & 102392 & kissat \\
    24 & BP\_4to2\_RC & 1.57 & 107072 & fraig \\
    25 & BP\_DT\_BK & 205.22 & 116008 & sat-sweeping \\
    26 & BP\_DT\_HCA & 374.82 & 109848 & kissat \\
    27 & BP\_DT\_JCA & 357.18 & 111816 & kissat \\
    28 & BP\_DT\_KS & 672.73 & 115396 & kissat \\
    29 & BP\_DT\_LF & 348.69 & 111860 & kissat \\
    30 & BP\_DT\_RC & 0.88 & 105508 & fraig \\
    31 & BP\_WT\_BK & 237.07 & 108100 & kissat \\
    32 & BP\_WT\_HCA & 382.67 & 108456 & kissat \\
    \bottomrule
  \end{tabular}
\end{table}

\begin{table}[H]
  \centering
  \small
  \setlength{\tabcolsep}{4pt}
  \renewcommand{\arraystretch}{1.2}
  \begin{tabular}{@{}r l l l l@{}}
    \toprule
    \multicolumn{1}{c}{\#} & \multicolumn{1}{c}{Tested\_resyn3\_128\_circuit} & \multicolumn{1}{c}{Time (seconds)} & \multicolumn{1}{c}{Memory (KB)} & \multicolumn{1}{c}{Method} \\
    \midrule
    33 & BP\_WT\_JCA & 645.68 & 112884 & kissat \\
    34 & BP\_WT\_KS & 667.53 & 111092 & kissat \\
    35 & BP\_WT\_LF & 637.84 & 108664 & kissat \\
    36 & BP\_WT\_RC & 269.41 & 110884 & sat-sweeping \\
    37 & SP\_AR\_BK & 43.62 & 76176 & sat-sweeping \\
    38 & SP\_AR\_CK & 6.14 & 130784 & fraig \\
    39 & SP\_AR\_CL & 4947.87 & 276568 & sat-sweeping \\
    40 & SP\_AR\_KS & 1257.36 & 233016 & sat-sweeping \\
    41 & SP\_AR\_LF & 461.84 & 238684 & sat-sweeping \\
    42 & SP\_AR\_RC & 5.01 & 126128 & fraig \\
    43 & SP\_AR\_SE & 4.54 & 129400 & fraig \\
    44 & SP\_CWT\_BK & 115.41 & 114264 & kissat \\
    45 & SP\_CWT\_CK & 37.21 & 126924 & sat-sweeping \\
    46 & SP\_CWT\_CL & 3524.06 & 186608 & kissat \\
    47 & SP\_CWT\_KS & 240.08 & 88308 & sat-sweeping \\
    48 & SP\_CWT\_LF & 116.71 & 124204 & kissat \\
    49 & SP\_CWT\_RC & 27.63 & 92696 & sat-sweeping \\
    50 & SP\_CWT\_SE & 1.07 & 117224 & fraig \\
    51 & SP\_DT\_BK & 39.06 & 106960 & kissat \\
    52 & SP\_DT\_CK & 28.19 & 66628 & sat-sweeping \\
    53 & SP\_DT\_CL & 2450.09 & 149456 & kissat \\
    54 & SP\_DT\_KS & 122.68 & 108596 & kissat \\
    55 & SP\_DT\_LF & 71.81 & 107388 & kissat \\
    56 & SP\_DT\_RC & 1.20 & 113388 & fraig \\
    57 & SP\_DT\_SE & 0.76 & 81376 & fraig \\
    58 & SP\_WT\_BK & 4.71 & 68092 & sat-sweeping \\
    59 & SP\_WT\_CK & 59.40 & 108668 & kissat \\
    60 & SP\_WT\_CL & 1020.57 & 127684 & kissat \\
    61 & SP\_WT\_KS & 110.56 & 87728 & sat-sweeping \\
    62 & SP\_WT\_LF & 78.41 & 114068 & kissat \\
    63 & SP\_WT\_RC & 40.93 & 94736 & sat-sweeping \\
    64 & SP\_WT\_SE & 1.36 & 116064 & fraig \\
    \bottomrule
  \end{tabular}
    \label{tab:test_resyn3_appendix}
\end{table}

\end{document}